\documentclass[twocolumn,showpacs,preprintnumbers,amsmath,amssymb,aps]{revtex4}

\usepackage{bm}
\usepackage{feynmf}
\usepackage[footnotesize,bf,centerlast]{caption}
\usepackage{hyperref}\hypersetup{colorlinks=true,linkcolor=blue,citecolor=red,urlcolor=magenta,bookmarksnumbered=true,bookmarksopen=true,bookmarksopenlevel=1}
\usepackage{graphicx}
\usepackage{grffile}
\usepackage{subfigure}

\renewcommand{\d}{\mathrm{d}}
\newcommand{\dalembert}{{\scriptstyle \square}}
\newcommand{\ea}{\textit{et al}.\ }
\newcommand{\eq}[1]{(\ref{#1})}
\newcommand{\feynd}[1]{#1\kern-0.55em/}
\newcommand{\feynk}[1]{#1\kern-0.50em/}
\newcommand{\feynp}[1]{#1\kern-0.45em/}
\newcommand{\feynP}[1]{#1\kern-0.65em/}
\newcommand{\feynS}[1]{#1\kern-0.60em/}

\renewcommand{\L}{\mathcal{L}}
\renewcommand{\O}{\mathcal{O}}
\renewcommand{\P}{\mathcal{P}}
\newcommand{\p}{\partial}

\begin{document}

\title{Consistent interactions for high-spin fermion fields}

\author{Tom Vrancx}
\email{Tom.Vrancx@UGent.be}
\author{Lesley De Cruz}
\author{Jan Ryckebusch}
\author{Pieter Vancraeyveld}

\affiliation{Department of Physics and Astronomy,\\
 Ghent University, Proeftuinstraat 86, B-9000 Gent, Belgium}
\date{\today}

\begin{abstract}
We address the issue of consistent interactions for off-shell fermion fields of arbitrary spin. These interactions play a crucial role in the quantum hadrodynamical description of high-spin baryon resonances in hadronic processes. The Rarita-Schwinger description of high-spin fermion fields involves unphysical degrees of freedom, associated with their lower-spin content. These enter the interaction if not eliminated outright. The invariance condition of the interaction under the unconstrained Rarita-Schwinger gauge removes the lower-spin content of the fermion propagator and leads to a consistent description of the interaction. 
We develop the most general, consistent interaction structure for high-spin fermions. We find that the power of the momentum dependence of a consistent interaction rises with the spin of the fermion field. This leads to unphysical structures in the energy dependence of the computed tree-level cross sections when the short-distance physics is cut off with standard hadronic form factors. A novel, spin-dependent hadronic form factor is proposed that suppresses the unphysical artifacts.
\end{abstract}



\maketitle

\section{Introduction}
\label{sec:introduction}
In 1941, William Rarita and Julian Schwinger proposed a quantum field theory for particles with an arbitrary half integral spin $s$ \cite{Rarita:1941mf}. Today, 70 years later, this theory is commonly used to describe phenomena that involve relativistic high-spin ($s \ge {3/2}$) fermions. In the Rarita-Schwinger (R-S) theory a relativistic spin-$s$ fermion, with $s = n + {1/2}$ and $n \in \mathbb{N}^*$, is described by the totally symmetric spin-$\left(n + {1/2}\right)$ field $\psi_{\mu_1\ldots\mu_n}$, which obeys the so-called R-S equations
\begin{subequations}
\label{RS-equations}
\begin{align}
(i\feynd{\p} - m)\psi_{\mu_1\ldots\mu_n} &= 0, \label{RS-equations-1}\\
\gamma^{\mu_1}\psi_{\mu_1\mu_2\ldots\mu_n} &= 0 \label{RS-equations-2}.
\end{align}
\end{subequations}
These equations comprise the equations of motion \eq{RS-equations-1}, which are akin to those of the Dirac theory, and the R-S constraints \eq{RS-equations-2}. By construction, R-S fields have redundant degrees of freedom (dof), which are associated with unphysical lower-spin fields. Massive R-S fields have $\frac{4(n+3)!}{3!n!}$ dof whereas the required number is $4(n+1)$. The R-S constraints \eq{RS-equations-2} eliminate the redundant dof of non-interacting R-S fields \cite{Rarita:1941mf}. Massless R-S fields, however, can have only four dof in order to guarantee consistency with the special theory of relativity. Therefore, the massless R-S theory should be invariant under the spin-$(n+1/2)$ R-S gauge (RS$_{n+1/2}$) \cite{Pascalutsa:1999zz}
\begin{align}
&\psi_{\mu_1\ldots\mu_n} \rightarrow \psi_{\mu_1\ldots\mu_n} + \frac{i}{n(n-1)!}\sum_{P(\mu)}\p_{\mu_1}\xi_{\mu_2\ldots\mu_n}, \nonumber\\
&\gamma^{\mu_1}\xi_{\mu_1\mu_2\ldots\mu_{n-1}} = 0. \label{RS-gauge}
\end{align}
The notation $\sum_{P(\mu)}$ denotes the summing over all the permutations of the $\mu_i$ indices. Further, $\xi_{\mu_1\ldots\mu_{n-1}}$ represents a totally symmetric, space-time-dependent rank-$(n-1)$ tensor-spinor field. The inclusion of the factor $i$ in Eq.~\eq{RS-gauge} is a convention. If $\psi_{\mu_1\ldots\mu_n}$ is a real field, then the field $\xi_{\mu_1\ldots\mu_{n-1}}$ has to be defined as an imaginary field.

The interacting case is more convoluted. If the R-S field is off its mass shell (``off-shell'') the unphysical R-S components might participate in the interaction if not eliminated in a controlled way. Such an interaction is dubbed ``inconsistent'' since it is not mediated purely by the physical component of the R-S field. Several, physically unacceptable, problems are associated with these inconsistent interactions, the most famous of which are the Johnson-Sudarshan and Velo-Zwanzinger problems \cite{Johnson:1960vt,Velo:1969bt}. 

In 1998, Vladimir Pascalutsa succeeded in formulating a consistent interaction theory for massive spin-${3/2}$ fields \cite{Pascalutsa:1998pw}. The consistency is provided by the invariance of the spin-${3/2}$ interaction under the local $U(1)$ gauge. In Sec.~\ref{subsec:spin-52} of this paper Pascalutsa's theory is extended and a consistent interaction theory for massive spin-${5/2}$ fields is developed. In Sec.~\ref{subsec:spin-J}, the most general consistent interaction structure for massive spin-$\left(n + {1/2}\right)$ fields is derived, based on a generalization of the results obtained for the spin-${5/2}$ theory in Sec.~\ref{subsec:spin-52}. Then, consistent couplings for the $(\phi\psi\psi^*_{\mu_1\ldots\mu_n})$- and $(A_\mu\psi\psi^*_{\mu_1\ldots\mu_n})$-theories are constructed. Here, the fields $\phi$, $\psi$ and $A_\mu$ represent a spin-$0$ field, a spin-${1/2}$ field and a spin-$1$ field. Section \ref{sec:hadron_physics} illustrates the application of the consistent $(\phi\psi\psi^*_{\mu_1\ldots\mu_n})$- and $(A_\mu\psi\psi^*_{\mu_1\ldots\mu_n})$-couplings in hadron physics. In Sec.~\ref{subsec:standard_HFF}, the problems which arise when combining consistent high-spin interactions with standard hadronic form factors are discussed. To our knowledge, this issue has never been pointed out in the relevant literature so far. An alternative hadronic form factor is proposed in Sec.~\ref{subsec:multidipole-Gauss_form_factor} which remedies the issues mentioned in Sec.~\ref{subsec:standard_HFF}. Finally, Sec.~\ref{sec:conclusions} states the conclusions of this work.

\section{Consistent interactions}
\label{sec:consistent_interactions}
\subsection{The massive spin-${5/2}$ field}
\label{subsec:spin-52}
\subsubsection{Gauge-invariance as a requirement for a consistent interaction theory}
\label{subsubsec:gauge-invariance}
A general interaction theory for a massive off-shell spin-${5/2}$ field $\psi_{\mu\nu}$ and an on-shell source $J_{\mu\nu}$ can be constructed from a Lagrangian of the type
\begin{align}
\L_I = \overline{\psi}_{\mu\nu}J^{\mu\nu} + \overline{J}_{\mu\nu}\psi^{\mu\nu}, \label{spin-52_interaction}
\end{align}
where the subscript $I$ stands for ``interaction''. In order to obtain a consistent spin-$5/2$ interaction theory, $\L_I$ has to be constructed in such a way that only the physical $\P^{{5/2}}_{\mu\nu;\lambda\rho}(\p)\psi^{\lambda\rho}$ component of $\psi_{\mu\nu}$ mediates the interaction, i.e.\
\begin{align}
&\overline{\Gamma}_f^{\mu\nu}(\{p_f\})P_{\mu\nu;\lambda\rho}(p)\Gamma_i^{\lambda\rho}(\{p_i\}) \nonumber\\
&= \overline{\Gamma}_f^{\mu\nu}(\{p_f\})\frac{\feynp{p} + m}{p^2 - m^2}\P^{{5/2}}_{\mu\nu;\lambda\rho}(p)\Gamma_i^{\lambda\rho}(\{p_i\}),\label{52-consistent-interaction}
\end{align}
with $m$ the mass of $\psi_{\mu\nu}$, $\Gamma_{i,f}^{\mu\nu}(\{p_{i,f}\})$ the initial/final interaction vertex which is derived from the source $J_{i,f}^{\mu\nu}$, $P_{\mu\nu;\lambda\rho}(p)$ the spin-${5/2}$ propagator and $\P^{{5/2}}_{\mu\nu;\lambda\rho}(p)$ the spin-${5/2}$ projection operator. Further, $p$ represents the four-momentum of $\psi_{\mu\nu}$ and $\{p_{i,f}\}$ denotes the collection of four-momenta of the fields that are contained in $J_{i,f}^{\mu\nu}$. The explicit expressions for the spin-${5/2}$ projection operator and the spin-${3/2}$ and spin-${1/2}$ projection operators, which project the field $\psi_{\mu\nu}$ onto its unphysical spin-${3/2}$ and spin-${1/2}$ components, are given in Appendix \ref{appendix:projection_operators}. 

The propagator $P_{\mu\nu;\lambda\rho}(p)$ can be projected onto the spin-${5/2}$, spin-${3/2}$ and spin-${1/2}$ projection operators, which span the complete spin-$5/2$ space in the R-S representation. As is seen from Eqs.~\eq{P52}, \eq{P32} and \eq{P12}, the spin projection operators contain terms that are proportional to $p^{-2}$ and $p^{-4}$. However, in order to describe a massive off-shell spin-$5/2$ field in a physically meaningful way, $P_{\mu\nu;\lambda\rho}(p)$ has to be regular for $p^2 \rightarrow 0$. So, in the expression for $P_{\mu\nu;\lambda\rho}(p)$ the singular terms, stemming from the spin projection operators, have to cancel each other out. 

In a consistent spin-$5/2$ interaction only the physical spin-$5/2$ component of $P_{\mu\nu;\lambda\rho}(p)$ remains, as is expressed through Eq.~\eq{52-consistent-interaction}. Since the left-hand side of Eq.~\eq{52-consistent-interaction} is regular for $p^2\rightarrow 0$, the right-hand side of this equation has to be regular for $p^2\rightarrow 0$ as well. So, in order for Eq.~\eq{52-consistent-interaction} to hold, the singular terms of the spin-$5/2$ projection operator have to be removed by the interaction vertices $\Gamma_{i,f}^{\mu\nu}(\{p_{i,f}\})$. This implies that the interaction vertices cannot be completely arbitrary. Instead they have to be constrained by a certain local symmetry. This local symmetry can readily be found. Indeed, all of the singular terms of $\P^{{5/2}}_{\mu\nu;\lambda\rho}(p)$ are at least linear in $p_\mu$, $p_\nu$, $p_\lambda$ or $p_\rho$. It is straightforward to show that the right-hand side of Eq.~\eq{52-consistent-interaction} is only regular for $p^2\rightarrow 0$ when the interaction vertices satisfy the following transversality conditions
\begin{align}
p_\mu \Gamma_I^{\mu\nu} &= 0, \nonumber\\
p_\nu \Gamma_I^{\mu\nu} &=0.\label{transverse_vertices}
\end{align}
This requirement is equivalent to the invariance of $\L_I$ under the unconstrained spin-$5/2$ R-S (uRS$_{5/2}$) gauge
\begin{align}
\psi_{\mu\nu} \rightarrow \psi_{\mu\nu} + \frac{i}{2}\left(\p_\mu\chi_\nu + \p_\nu\chi_\mu\right), \label{RS-gauge-2}
\end{align}
where $\chi_\mu$ represents an arbitrary, space-time-dependent rank-one tensor-spinor field. The invariance of interaction theories under the uRS$_{5/2}$ gauge of Eq.~\eq{RS-gauge-2} is the required local symmetry that guarantees the consistency of massive, off-shell spin-$5/2$ interactions.

%

Constructing interaction vertices (which are derived from the sources) that satisfy conditions \eq{transverse_vertices} is not a trivial task. However, by using a field construction that is invariant under the uRS$_{5/2}$ gauge \eq{RS-gauge-2}, the problem of finding conserved sources can be circumvented.

\subsubsection{The gauge-invariant field}
\label{subsubsec:gauge-invariant_field}
Inspired by the gauge-invariant field
\begin{align}
G_{\mu,\nu} = i\left(\p_\mu\psi_\nu - \p_\nu\psi_\mu\right),
\end{align}
which was introduced by Pascalutsa in Ref.~\cite{Pascalutsa:1998pw} to set up consistent interaction theories for spin-${3/2}$ particles, a gauge-invariant spin-${5/2}$ field can be constructed. The field $G_{\mu,\nu}$ can be rewritten as
\begin{align}
G_{\mu,\nu} &= i\left(\p_\mu g_{\nu\lambda} - \p_\nu g_{\mu\lambda}\right)\psi^\lambda, \nonumber\\
&= O^{{3/2}}_{(\mu,\nu)\lambda}(\p)\psi^\lambda,
\end{align}
which reveals the interaction operator $O^{{3/2}}_{(\mu,\nu)\lambda}(\p) = -O^{{3/2}}_{(\nu,\mu)\lambda}(\p)$. This operator has the following property
\begin{align}
\p^\lambda O^{{3/2}}_{(\mu,\nu)\lambda}(\p) = O^{{3/2}}_{(\mu,\nu)\lambda}(\p)\p^\lambda = 0,
\end{align}
which ensures the invariance of $G_{\mu,\nu}$ under the uRS$_{5/2}$ gauge. The notation ``$(\mu,\nu)\lambda$'' for the tensor indices of the spin-${3/2}$ interaction operator is used to separate the actual Lorentz indices of the gauge-invariant field, i.e.\ $\mu$ and $\nu$, from the Lorentz index that is contracted with the spin-${3/2}$ field, i.e.\ $\lambda$. The corresponding spin-${5/2}$ interaction operator, i.e.\ $O^{{5/2}}_{(\mu\nu,\lambda\rho)\sigma\tau}(\p)$, is constructed from the direct product of two spin-${3/2}$ interaction operators
\begin{align}
&O^{{5/2}}_{(\mu\nu,\lambda\rho)\sigma\tau}(\p) \nonumber \\
&= \frac{1}{4}\Bigl(O^{{3/2}}_{(\mu,\lambda)\sigma}(\p)  O^{{3/2}}_{(\nu,\rho)\tau}(\p) + O^{{3/2}}_{(\mu,\rho)\sigma}(\p)  O^{{3/2}}_{(\nu,\lambda)\tau}(\p) \nonumber \\
&\quad\,+ O^{{3/2}}_{(\mu,\lambda)\tau}(\p)  O^{{3/2}}_{(\nu,\rho)\sigma}(\p) + O^{{3/2}}_{(\mu,\rho)\tau}(\p)  O^{{3/2}}_{(\nu,\lambda)\sigma}(\p)\Bigr), \nonumber\\
&= -\frac{1}{2}\p_\mu\p_\nu \left(g_{\lambda\sigma}g_{\rho\tau} + g_{\lambda\tau}g_{\rho\sigma}\right)  \nonumber\\
&\quad\,- \frac{1}{2}\p_\lambda\p_\rho \left(g_{\mu\sigma}g_{\nu\tau} + g_{\mu\tau}g_{\nu\sigma}\right) \nonumber\\ 
&\quad\,+ \frac{1}{4}\p_\mu\p_\lambda \left(g_{\nu\sigma}g_{\rho\tau} + g_{\nu\tau}g_{\rho\sigma}\right) \nonumber\\
&\quad\,+ \frac{1}{4}\p_\mu\p_\rho \left(g_{\nu\sigma}g_{\lambda\tau} + g_{\nu\tau}g_{\lambda\sigma}\right) \nonumber\\
&\quad\,+ \frac{1}{4}\p_\nu\p_\lambda \left(g_{\mu\sigma}g_{\rho\tau} + g_{\mu\tau}g_{\rho\sigma}\right) \nonumber\\
&\quad\,+ \frac{1}{4}\p_\nu\p_\rho \left(g_{\mu\sigma}g_{\lambda\tau} + g_{\mu\tau}g_{\lambda\sigma}\right)\label{O52}.
\end{align}
The corresponding gauge-invariant field for the spin-${5/2}$ theory reads
\begin{align}
G_{\mu\nu,\lambda\rho} &= O^{{5/2}}_{(\mu\nu,\lambda\rho)\sigma\tau}(\p)\psi^{\sigma\tau}, \nonumber\\
&= -\p_\mu\p_\nu\psi_{\lambda\rho} - \p_\lambda\p_\rho\psi_{\mu\nu}  \nonumber\\
&\quad\,+ \frac{1}{2}(\p_\mu\p_\lambda\psi_{\nu\rho} + \p_\mu\p_\rho\psi_{\nu\lambda} \nonumber\\
&\quad\,+ \p_\nu\p_\lambda\psi_{\mu\rho} + \p_\nu\p_\rho\psi_{\mu\lambda}),
\end{align}
where the symmetry of the spin-${5/2}$ field, i.e.\ $\psi_{\mu\nu} = \psi_{\nu\mu}$, has been applied. This field is invariant under the uRS$_{5/2}$ gauge \eq{RS-gauge-2} since
\begin{align}
\p^\sigma O^{{5/2}}_{(\mu\nu,\lambda\rho)\sigma\tau}(\p) &= O^{{5/2}}_{(\mu\nu,\lambda\rho)\sigma\tau}(\p)\p^\sigma = 0, \nonumber\\
\p^\tau O^{{5/2}}_{(\mu\nu,\lambda\rho)\sigma\tau}(\p) &= O^{{5/2}}_{(\mu\nu,\lambda\rho)\sigma\tau}(\p)\p^\tau = 0.\label{O52-properties}
\end{align}
The specific definition \eq{O52} for the spin-${5/2}$ interaction operator is chosen so as to symmetrize $G_{\mu\nu,\lambda\rho}$, i.e.\
\begin{align}
G_{\mu\nu,\lambda\rho} = G_{\nu\mu,\lambda\rho} &= G_{\mu\nu,\rho\lambda} = G_{\nu\mu,\rho\lambda}, \nonumber\\
G_{\mu\nu,\lambda\rho} &= G_{\lambda\rho,\mu\nu}.
\end{align}
An interaction theory constructed from $G_{\mu\nu,\lambda\rho}$ and a source $T_{\mu\nu\lambda\rho}$, i.e.\
\begin{align}
\L_I = \overline{G}_{\mu\nu,\lambda\rho}T^{\mu\nu\lambda\rho} + \overline{T}_{\mu\nu\lambda\rho}G^{\mu\nu,\lambda\rho}, \label{L_I}
\end{align}
consequently generates interaction vertices which obey the transversality conditions of Eq.~\eq{transverse_vertices}. 

\subsubsection{The propagator and the consistent interaction structure}
\label{subsubsec:propagator-interaction_structure}
Apart from the interaction vertices, the Feynman propagator is another key element in the quantum field theory of an interaction.
The Lagrangian $\L_I$, defined in Eq.~\eq{L_I}, gives rise to the following interaction structure
\begin{align}
O^{{5/2}}_{(\mu\nu,\lambda\rho)\sigma\tau}(p)P^{\sigma\tau;\sigma'\tau'}(p)O^{{5/2}}_{(\mu'\nu',\lambda'\rho')\sigma'\tau'}(p), \label{interaction_structure}
\end{align}
in the transition amplitude. In Ref.~\cite{Shklyar:2009cx}, Shklyar \ea derived an explicit form for the spin-$5/2$ propagator $P_{\mu\nu;\lambda\rho}(p)$. This propagator form, however, is not suitable for the current analysis because it does not lead to the consistent interaction structure of Eq.~\eq{52-consistent-interaction}. In other words, with the propagator of Ref.~\cite{Shklyar:2009cx}, the imposed condition of invariance of the interaction theory under the uRS$_{5/2}$ gauge \eq{RS-gauge-2} is not sufficient to remove the unphysical interactions from the Feynman amplitude. Indeed, using properties \eq{O52-properties} in momentum space and the explicit definitions of the lower-spin projection operators \eq{P32} and \eq{P12}, it is easy to show that
\begin{align}
&O^{{5/2}}_{(\mu\nu,\lambda\rho)\sigma\tau}(p)\P^{3/2,\sigma\tau;\sigma'\tau'}_{ij}(p)O^{{5/2}}_{(\mu'\nu',\lambda'\rho')\sigma'\tau'}(p) = 0, \nonumber\\
&i,j = 1,2 \qquad ij \neq 22,
\end{align}
and
\begin{align}
&O^{{5/2}}_{(\mu\nu,\lambda\rho)\sigma\tau}(p)\P^{1/2,\sigma\tau;\sigma'\tau'}_{kl}(p)O^{{5/2}}_{(\mu'\nu',\lambda'\rho')\sigma'\tau'}(p) = 0, \nonumber\\
&k,l = 1,2,3 \qquad kl \neq 22.
\end{align}
This implies that all the lower-spin components of $\psi_{\mu\nu}$, except for $\P^{{1/2}}_{22,\mu\nu;\lambda\rho}(\p)\psi^{\lambda\rho}$ and $\P^{{3/2}}_{22,\mu\nu;\lambda\rho}(\p)\psi^{\lambda\rho}$, decouple from a gauge-invariant spin-${5/2}$ interaction. The propagator from Ref.~\cite{Shklyar:2009cx}, however, contains these projection operators for arbitrary values of its two free parameters. For a specific choice of one of the parameters, $\P^{{3/2}}_{22,\mu\nu;\lambda\rho}(p)$ can be removed from the propagator. However, this is not the case for the spin-${1/2}$ sector and  the interaction will always receive contributions from the propagating $\P^{{1/2}}_{22,\mu\nu;\lambda\rho}(\p)\psi^{\lambda\rho}$ component. This peculiarity is intimately related to the infamous discontinuity in the R-S description of spin-${5/2}$ fields, which was found by Berends \ea in Ref.~\cite{Berends:1979rv}: the zero-mass limit of the massive theory does not coincide with the massless theory, which is invariant under the RS$_{5/2}$ gauge \eq{RS-gauge}. The conclusion is that a ``gauge-invariant'' spin-${5/2}$ propagator, i.e.\ without terms proportional to $\P^{{1/2}}_{22,\mu\nu;\lambda\rho}(p)$ and $\P^{{3/2}}_{22,\mu\nu;\lambda\rho}(p)$, cannot be derived from the massive spin-${5/2}$ theory. It is clear that in order to construct a consistent interaction from a gauge-invariant theory, a different approach should be adopted.

The commonly used spin-${3/2}$ propagator reads \cite{Benmerrouche:1989uc}
\begin{align}
P_{\mu;\nu}(p) &= \frac{\feynp{p} + m}{p^2 - m^2}\biggl(g_{\mu\nu} - \frac{1}{3}\gamma_\mu\gamma_\nu  \nonumber\\
&\quad\,- \frac{1}{3m}(\gamma_\mu p_\nu - \gamma_\nu p_\mu) -\frac{2}{3m^2}p_\mu p_\nu\biggr).\label{propagator-32}
\end{align}
It is important to note that this propagator results from a massive theory that is invariant under the RS$_{3/2}$ gauge in the zero-mass limit \cite{Pascalutsa:1998pw}. The projection operator that projects the spin-${3/2}$ field $\psi_\mu$ onto the physical spin-${3/2}$ component $\P^{{3/2}}_{\mu\nu}(\p)\psi^\nu$ is given by \cite{Benmerrouche:1989uc}
\begin{align}
\P^{{3/2}}_{\mu;\nu}(p) &= \biggl(g_{\mu\nu} - \frac{1}{3}\gamma_\mu \gamma_\nu \nonumber\\
&\quad\,- \frac{\feynp{p}}{3p^2}(\gamma_\mu p_\nu - \gamma_\nu p_\mu) - \frac{2}{3p^2} p_\mu p_\nu \biggr).\label{projection-32}
\end{align}
A closer inspection shows that the ``gauge-invariant'' spin-${3/2}$ propagator \eq{propagator-32} can be obtained from the spin-${3/2}$ projection operator \eq{projection-32} by means of the following substitutions
\begin{align}
\feynp{p} &\rightarrow m, \nonumber\\
p^2 &\rightarrow m^2,\label{P52_to_P}
\end{align}
and subsequently multiplying it with $(\feynp{p}-m)^{-1}$. Since $\left[\feynp{p},\gamma_\mu\right] \neq 0$, the subsitution $\feynp{p} \rightarrow m$ of Eq.~\eq{P52_to_P} only holds when all the $\feynp{p}$'s are moved to the left from the Dirac matrices, which is the case in expression \eq{projection-32}. Equivalently, when all the $\feynp{p}$'s are moved to the right from the Dirac matrices the substitution $\feynp{p} \rightarrow -m$ should be carried out. Applying the prescription of Eq.~\eq{P52_to_P} to the spin-${5/2}$ projection operator \eq{P52} results in the following expression for the spin-${5/2}$ propagator
\begin{align}
&P_{\mu\nu;\lambda\rho}(p) \nonumber\\
&= \frac{\feynp{p} + m}{p^2 - m^2}\biggl[\frac{1}{2}(g_{\mu\lambda}g_{\nu\rho}+g_{\mu\rho}g_{\nu\lambda}) - \frac{1}{5}g_{\mu\nu}g_{\lambda\rho} \nonumber\\
&\quad\,- \frac{1}{10}(g_{\mu\lambda}\gamma_\nu\gamma_\rho + g_{\mu\rho}\gamma_\nu\gamma_\lambda + g_{\nu\lambda}\gamma_\mu\gamma_\rho + g_{\nu\rho}\gamma_\mu\gamma_\lambda) \nonumber\\
&\quad\,+ \frac{1}{10m}\Bigl(g_{\mu\lambda}(p_\nu\gamma_\rho - p_\rho\gamma_\nu) + g_{\mu\rho}(p_\nu\gamma_\lambda - p_\lambda\gamma_\nu)  \nonumber\\
&\quad\,+ g_{\nu\lambda}(p_\mu\gamma_\rho - p_\rho\gamma_\mu) + g_{\nu\rho}(p_\mu\gamma_\lambda - p_\lambda\gamma_\mu)\Bigr) \nonumber\\
&\quad\,+ \frac{1}{5m^2}(g_{\mu\nu}p_\nu p_\rho + g_{\lambda\rho}p_\mu p_\nu) \nonumber\\
&\quad\,- \frac{2}{5m^2}(g_{\mu\lambda}p_\nu p_\rho + g_{\mu\rho}p_\nu p_\lambda + g_{\nu\lambda}p_\mu p_\rho + g_{\nu\rho}p_\mu p_\lambda) \nonumber\\
&\quad\,+ \frac{1}{10m^2}(\gamma_\mu\gamma_\lambda p_\nu p_\rho + \gamma_\mu\gamma_\rho p_\nu p_\lambda \nonumber\\
&\quad\,+ \gamma_\nu\gamma_\lambda p_\mu p_\rho + \gamma_\nu\gamma_\rho p_\mu p_\lambda) \nonumber\\
&\quad\,+ \frac{1}{5m^3}(\gamma_\mu p_\nu p_\lambda p_\rho + \gamma_\nu p_\mu p_\lambda p_\rho \nonumber\\
&\quad\,- \gamma_\lambda p_\mu p_\nu p_\rho - \gamma_\rho p_\mu p_\nu p_\lambda) + \frac{2}{5m^4}p_\mu p_\nu p_\lambda p_\rho\biggr]\label{propagator-52-1}.
\end{align}
Expanding $P_{\mu\nu;\lambda\rho}(p)$ in terms of the projection operators defined in Eqs.~\eq{P52}, \eq{P32} and \eq{P12} leads to the following expression
\begin{align}
&P_{\mu\nu;\lambda\rho}(p) \nonumber\\
&= \biggl[\frac{\feynp{p} + m}{p^2 - m^2}\P^{{5/2}} \nonumber\\
&\quad\,- \frac{4}{5m^2}(\feynp{p}+m)\P^{{3/2}}_{11} + \frac{1}{\sqrt{5}m}\left(\P^{{3/2}}_{12} + \P^{{3/2}}_{21}\right) \nonumber\\
&\quad\,+ \frac{2}{5m^4}(\feynp{p}+m)(p^2-m^2)\P^{{1/2}}_{11} - \frac{1}{5m^2}(\feynp{p}+m)\P^{{1/2}}_{33} \nonumber\\
&\quad\,+ \frac{\sqrt{3}}{5m^2}(\feynp{p}+m)\left(\P^{{1/2}}_{12} + \P^{{1/2}}_{21}\right) \nonumber\\
&\quad\,- \frac{\sqrt{6}}{5m^3}(p^2-m^2)\left(\P^{{1/2}}_{13} + \P^{{1/2}}_{31}\right) \nonumber\\
&\quad\,- \frac{\sqrt{2}}{5m}\left(\P^{{3/2}}_{32} + \P^{{3/2}}_{23}\right)\biggr]_{\mu\nu;\lambda\rho}(p)\label{propagator-52-2}. 
\end{align}
This propagator is a Hermitian operator
\begin{align}
P^\dagger_{\mu\nu;\lambda\rho}(p) &= \gamma_0P_{\lambda\rho;\mu\nu}(p)\gamma_0,\label{P52-Hermitian}
\end{align}
and has the following symmetry properties
\begin{align}
P_{\mu\nu;\lambda\rho}(p) = P_{\nu\mu;\lambda\rho}(p) = P_{\mu\nu;\rho\lambda}(p) = P_{\nu\mu;\rho\lambda}(p).\label{P52-symmetries}
\end{align}
Moreover, this propagator does not recieve contributions from $\P^{{1/2}}_{22,\mu\nu;\lambda\rho}(p)$ and $\P^{{3/2}}_{22,\mu\nu;\lambda\rho}(p)$. As a result, it generates consistent couplings from gauge-invariant interaction theories
\begin{align}
&O^{{5/2}}_{(\mu\nu,\lambda\rho)\sigma\tau}(p)P^{\sigma\tau;\sigma'\tau'}(p)O^{{5/2}}_{(\mu'\nu',\lambda'\rho')\sigma'\tau'}(p) \nonumber\\
&=\frac{\feynp{p} + m}{p^2 - m^2}O^{{5/2}}_{(\mu\nu,\lambda\rho)\sigma\tau}(p)\P^{5/2,\sigma\tau;\sigma'\tau'}(p)O^{{5/2}}_{(\mu'\nu',\lambda'\rho')\sigma'\tau'}(p),\label{OPO-consistent_interactions}
\end{align}
which is precisely of the form proposed in Eq.~\eq{52-consistent-interaction}. As a result, the unphysical components in the propagator completely decouple from the interaction which is fully mediated by the physical spin-$5/2$ component $\P^{{5/2}}_{\mu\nu;\lambda\rho}(\p)\psi^{\lambda\rho}$. Combining Eqs.~\eq{O52-properties} and \eq{propagator-52-1}, the consistent interaction structure \eq{OPO-consistent_interactions} can be reduced to
\begin{align}
&O^{{5/2}}_{(\mu\nu,\lambda\rho)\sigma\tau}(p)P^{\sigma\tau;\sigma'\tau'}(p)O^{{5/2}}_{(\mu'\nu',\lambda'\rho')\sigma'\tau'}(p) \nonumber\\
&= \frac{\feynp{p} + m}{p^2 - m^2}O^{{5/2}}_{(\mu\nu,\lambda\rho)\sigma\tau}(p)O^{{5/2}}_{(\mu'\nu',\lambda'\rho')\sigma'\tau'}(p) \nonumber\\
&\quad\,\times\biggl[\frac{1}{2}\left(g^{\sigma{\sigma'}}g^{\tau{\tau'}}+g^{\sigma{\tau'}}g^{\tau{\sigma'}}\right) - \frac{1}{5}g^{\sigma\tau}g^{{\sigma'}{\tau'}} \nonumber\\
&\quad\,- \frac{1}{10}\Bigl(g^{\sigma{\sigma'}}\gamma^\tau\gamma^{\tau'} + g^{\sigma{\tau'}}\gamma^\tau\gamma^{\sigma'} \nonumber\\
&\quad\,+ g^{\tau{\sigma'}}\gamma^\sigma\gamma^{\tau'} + g^{\tau{\tau'}}\gamma^\sigma\gamma^{\sigma'}\Bigr)\biggr].\label{OPO-reduced}
\end{align}
From Eq.~\eq{OPO-reduced} it becomes clear why relation \eq{OPO-consistent_interactions} holds, without resorting to the decomposition \eq{propagator-52-2} of the propagator. The propagator was obtained from the spin-${5/2}$ projection operator by means of the substitutions \eq{P52_to_P}. These only affect the terms that are at least linear in $p_\mu$, which all vanish upon contraction with the interaction operator since relations \eq{O52-properties} hold. As a result, only the physical component of the propagator remains present in \eq{OPO-consistent_interactions}.

The above-mentioned method for deriving $P_{\mu\nu;\lambda\rho}(p)$ raises some issues. As a matter of fact, there is an infinite number of propagators that satisfy conditions \eq{P52-Hermitian} and \eq{P52-symmetries} and concurrently lead to consistent interactions. However, it is confirmed by Huang \ea in their work on the Feynman propagator for a particle with arbitrary spin \cite{Huang:2005js}, that expression \eq{propagator-52-1} is indeed the appropriate one for the spin-${5/2}$ propagator apart from a series of non-covariant terms. As emphasized by Steven Weinberg in Ref.~\cite{Weinberg:1964cn}, the non-covariant terms should be removed from the Feynman propagator in order to guarantee that the $S$-matrix remains invariant under proper orthochronous Lorentz transformations. The derivation of the Feynman propagator by Huang \ea does not rely on the expression for the free R-S Lagrangian. Instead the propagator is derived from its definition: the vacuum expectation value of the time-ordered product of the free R-S field and its adjoint. For the spin-${5/2}$ theory, this definition gives rise to the following propagator
\begin{align}
\Delta_F^{\mu\nu;\lambda\rho}(x-x') &= \langle 0 | \mathcal{T}\{\psi^{\mu\nu}(x)\overline{\psi}^{\lambda\rho}(x') \} | 0 \rangle, \nonumber\\
&= \theta(t-t')\langle 0 | \psi^{\mu\nu}(x)\overline{\psi}^{\lambda\rho}(x') | 0 \rangle \nonumber\\
&\quad\,- \theta(t'-t)\langle 0 |\overline{\psi}^{\lambda\rho}(x') \psi^{\mu\nu}(x) | 0 \rangle.
\end{align}
The calculation of the Feynman propagator by Huang \ea is based on the solutions to the R-S Eqs.~\eq{RS-equations}. Therefore, this method does not suffer from the mentioned discontinuity between the massive and the massless spin-${5/2}$ theory.

We mention that in Ref.~\cite{Shklyar:2004dy}, Shklyar \ea obtained a consistent spin-$5/2$ interaction by the explicit replacement
\begin{align}
P'_{\mu\nu;\lambda\rho}(p) \rightarrow \frac{p^4}{m^4}\frac{\feynp{p} + m}{p^2 - m^2} \P^{{5/2}}_{\mu\nu;\lambda\rho}(p), \label{Shklyar_substitution}
\end{align}
in the expression for the spin-$5/2$ interaction structure. They use the following expression for $P'_{\mu\nu;\lambda\rho}(p)$
\begin{align}
P'_{\mu\nu;\lambda\rho}(p) &= \frac{\feynp{p} + m}{p^2 - m^2}\biggl(\frac{1}{2}\left( S_{\mu\lambda} S_{\nu\rho} +  S_{\mu\rho} S_{\nu\lambda}\right) - \frac{1}{5} S_{\mu\nu} S_{\lambda\rho} \nonumber\\
&\quad\,+ \frac{1}{10}\bigl(\feynS{ S}_\mu\feynS{ S}_\lambda S_{\nu\rho} + \feynS{ S}_\mu\feynS{ S}_\rho S_{\nu\lambda} \nonumber\\
&\quad\,+ \feynS{ S}_\nu\feynS{ S}_\lambda S_{\mu\lambda} + \feynS{ S}_\nu\feynS{ S}_\rho S_{\mu\rho}\bigr)\biggr),
\end{align}
with
\begin{align}
S_{\mu\nu}(p) &= -g_{\mu\nu} + \frac{1}{m^2}p_\mu p_\nu,\nonumber\\
\feynS{S}_\mu(p) &= \gamma^\nu S_{\mu\nu}(p),\nonumber\\
&= -\gamma_\mu + \frac{\feynp{p}}{m^2}p_\mu.
\end{align}
Note that $P'_{\mu\nu;\lambda\rho}(p)$ does not coincide with $P_{\mu\nu;\lambda\rho}(p)$ of Eq.~\eq{propagator-52-1}. In Ref.~\cite{Shklyar:2004dy}, Shklyar \ea assume that a uRS$_{5/2}$ gauge-invariant interaction leads to the substitution \eq{Shklyar_substitution} in the spin-$5/2$ interaction structure. However, they do not prove this statement. Furthermore, $P'_{\mu\nu;\lambda\rho}(p)$ is not a spin-$5/2$ propagator, it is a ``regular'' spin-$5/2$ projection operator that is multiplied with $(\feynp{p}-m)^{-1}$. By ``regular'' we mean that the singular $p^{-2}$ and $p^{-4}$ factors of the spin-$5/2$ projection operator are replaced with the factors $m^{-2}$ and $m^{-4}$ respectively.

\subsection{Massive fermion fields with arbitrary spin}
\label{subsec:spin-J}
\subsubsection{The general consistent interaction structure}
\label{subsubsec:general_strucure}
In the previous section it was demonstrated how consistent interaction structures can be constructed for off-shell spin-${5/2}$ fields. These results will now be generalized to off-shell spin-$s$ fermions, with $s = n + {1/2}$ and $n \in \mathbb{N}^*$.

The generalized interaction operator is defined as
\begin{align}
&O^{n+{1/2}}_{(\mu_1\ldots\mu_n,\nu_1\ldots\nu_n)\lambda_1\ldots\lambda_n}(\p) \nonumber\\
&= \frac{1}{(n!)^2}\sum_{P(\nu)}\sum_{P(\lambda)}O^{{3/2}}_{(\mu_1,\nu_1)\lambda_1}(\p)  \cdots  O^{{3/2}}_{(\mu_n,\nu_n)\lambda_n}(\p).\label{O-general}
\end{align}
Note that 
\begin{align}
&\widetilde{O}^{n+{1/2}}_{(\mu_1\ldots\mu_n,\nu_1\ldots\nu_n)\lambda_1\ldots\lambda_n}(\p) \nonumber\\
&= \frac{1}{(n!)^3}\sum_{P(\mu)}\sum_{P(\nu)}\sum_{P(\lambda)}O^{{3/2}}_{(\mu_1,\nu_1)\lambda_1}(\p)  \cdots  O^{{3/2}}_{(\mu_n,\nu_n)\lambda_n}(\p), \nonumber\\
&= 0,
\end{align}
since this operator is symmetric under $\mu_i \leftrightarrow \nu_j$ and $O^{{3/2}}_{(\mu,\nu)\lambda} = -O^{{3/2}}_{(\nu,\mu)\lambda}$. The associated gauge-invariant field for the spin-$\left(n+{1/2}\right)$ theory is given by
\begin{align}
G_{\mu_1\ldots\mu_n,\nu_1\ldots\nu_n} = O^{n+{1/2}}_{(\mu_1\ldots\mu_n,\nu_1\ldots\nu_n)\lambda_1\ldots\lambda_n}(\p)\psi^{\lambda_1\ldots\lambda_n},\label{G-general-0}
\end{align}
where $\psi_{\mu_1\ldots\mu_n}$ represents the spin-$\left(n+{1/2}\right)$ R-S field. Considering the expression for $O^{n+{1/2}}_{(\mu_1\ldots\mu_n,\nu_1\ldots\nu_n)\lambda_1\ldots\lambda_n}(\p)$ of Eq.~\eq{O-general} and the total symmetry of $\psi_{\mu_1\ldots\mu_n}$ in its Lorentz indices, the expression for $G_{\mu_1\ldots\mu_n,\nu_1\ldots\nu_n}$ of Eq.~\eq{G-general-0} is reduced to
\begin{align}
G_{\mu_1\ldots\mu_n,\nu_1\ldots\nu_n} &= \frac{1}{n!}\sum_{P(\nu)}O^{{3/2}}_{(\mu_1,\nu_1)\lambda_1}(\p)  \cdots  O^{{3/2}}_{(\mu_n,\nu_n)\lambda_n}(\p) \nonumber\\
&\quad\,\times\psi^{\lambda_1\ldots\lambda_n}.\label{G-general-1}
\end{align}
This expression for $G_{\mu_1\ldots\mu_n,\nu_1\ldots\nu_n}$ can be reformulated as
\begin{align}
G_{\mu_1\ldots\mu_n,\nu_1\ldots\nu_n} &= \sum_{P(\mu)}\sum_{P(\nu)}\sum_{k=0}^{n}\mathcal{G}^n_k \p_{\nu_1}\cdots\p_{\nu_k} \nonumber\\
&\quad\,\times\p_{\mu_{k+1}}\cdots\p_{\mu_n}\psi_{\mu_1\ldots\mu_k\nu_{k+1}\ldots\nu_n}, \label{G-general-2}
\end{align}
with
\begin{align}
\mathcal{G}^n_k = \frac{i^n(-1)^k}{n!k!(n-k)!}.
\end{align}
The field $G_{\mu_1\ldots\mu_n,\nu_1\ldots\nu_n}$ is totally symmetric in its $\mu_i$ and $\nu_j$ indices and has the following property
\begin{align}
G_{\mu_1\ldots\mu_n,\nu_1\ldots\nu_n} = (-1)^nG_{\nu_1\ldots\nu_n,\mu_1\ldots\mu_n}.
\end{align}
Further, it is invariant under the uRS$_{n+1/2}$ gauge
\begin{align}
\psi_{\mu_1\ldots\mu_n} \rightarrow \psi_{\mu_1\ldots\mu_n} + \frac{i}{n(n-1)!}\sum_{P(\mu)}\p_{\mu_1}\chi_{\mu_2\ldots\mu_n},\label{general-gauge}
\end{align}
with $\chi_{\mu_1\ldots\mu_{n-1}}$ an arbitrary, totally symmetric, space-time-dependent rank-$(n-1)$ tensor-spinor field. The invariance of $G_{\mu_1\ldots\mu_n,\nu_1\ldots\nu_n}$ under the uRS$_{n+1/2}$ gauge \eq{general-gauge} is assured by the following properties of the interaction operator
\begin{align}
&\p^{\lambda_k}O^{n+{1/2}}_{(\mu_1\ldots\mu_n,\nu_1\ldots\nu_n)\lambda_1\ldots\lambda_n}(\p) \nonumber\\
&= O^{n+{1/2}}_{(\mu_1\ldots\mu_n,\nu_1\ldots\nu_n)\lambda_1\ldots\lambda_n}(\p)\p^{\lambda_k}, \nonumber\\
&= 0,\label{O-general-properties}
\end{align}
where $k$ runs from $1$ to $n$. The interaction theory
\begin{align}
\L_I &= \overline{G}_{\mu_1\ldots\mu_n,\nu_1\ldots\nu_n}T^{\mu_1\ldots\mu_n\nu_1\ldots\nu_n} \nonumber\\
&\quad\,+ \overline{T}_{\mu_1\ldots\mu_n\nu_1\ldots\nu_n}G^{\mu_1\ldots\mu_n,\nu_1\ldots\nu_n}, \label{L_I-general}
\end{align}
which couples the gauge-invariant field $G_{\mu_1\ldots\mu_n,\nu_1\ldots\nu_n}$ to the external source $T_{\mu_1\ldots\mu_n\nu_1\ldots\nu_n}$, generates transverse interaction vertices, i.e.\
\begin{align}
p_{\mu_k}\Gamma_I^{\mu_1\ldots\mu_n} &= 0, \label{transverse_vertices-general}
\end{align}
where $k$ runs from $1$ to $n$, $p_\mu$ represents the four-momentum of $\psi_{\mu_1\ldots\mu_n}$ and $\Gamma_I^{\lambda_1\ldots\lambda_n}$ the interaction vertex derived from $O^{n+{1/2}\qquad\quad\lambda_1\ldots\lambda_n}_{(\mu_1\ldots\mu_n,\nu_1\ldots\nu_n)}(\p)T^{\mu_1\ldots\mu_n\nu_1\ldots\nu_n}$.

The next step consists of defining the spin-$\left(n+{1/2}\right)$ projection operators. The general expressions for the spin projection operators for bosons and fermions were first derived by Behrends and Fronsdal in Ref.~\cite{Behrends:1957}. The spin-$\left(n+{1/2}\right)$ projection operator is defined as
\begin{align}
&\P^{n+{1/2}}_{\mu_1\ldots\mu_n;\nu_1\ldots\nu_n}(p) \nonumber\\
&= \frac{n+1}{2n+3}\gamma^\mu \P^{n+1}_{\mu\mu_1\ldots\mu_n;\nu\nu_1\ldots\nu_n}(p)\gamma^\nu, \label{projection-general-1}
\end{align}
and depends on the spin-$(n+1)$ projection operator. The expression for the spin-$n$ projection operator reads
\begin{align}
&\P^{n}_{\mu_1\ldots\mu_n;\nu_1\ldots\nu_n}(p) \nonumber\\
&= \frac{1}{n!^2}\sum_{P(\mu)}\sum_{P(\nu)}\sum_{k=0}^{k_\textrm{max}}A_k^n\P_{\mu_1\mu_2}\P_{\nu_1\nu_2}\cdots\P_{\mu_{2k-1}\mu_{2k}}\P_{\nu_{2k-1}\nu_{2k}} \nonumber\\
&\quad\,\times\prod_{i=2k+1}^n\P_{\mu_i\nu_i},
\end{align}
with
\begin{align}
k_{\textrm{max}} = \begin{cases}
                     \frac{n}{2} & n\; \textrm{even}, \\
		     \frac{n-1}{2} & n\; \textrm{odd},
                     \end{cases}
\end{align}
and $\P_{\mu\nu}$
\begin{align}
\P_{\mu\nu} = g_{\mu\nu} - \frac{1}{p^2}p_\mu p_\nu.
\end{align}
The coefficients $A_k^n$ are given as
\begin{align}
A_k^n = \frac{1}{(-2)^k}\frac{n!}{k!(n-2k)!}\frac{(2n-2k-1)!!}{(2n-1)!!}. \label{Akn-coefficients}
\end{align}
However, expression \eq{projection-general-1} for the spin-$\left(n+{1/2}\right)$ projection operator can be further elaborated by explicitly carrying out the contractions with the Dirac matrices. This leads to a more convenient expression for the projection operator. Through a series of tedious calculations, which is the subject of Appendix \ref{sec:simplification_projection_operator}, the definition of the spin-$\left(n+{1/2}\right)$ projection operator is reformulated as
\begin{widetext}
\begin{align}
\P^{n+{1/2}}_{\mu_1\ldots\mu_n;\nu_1\ldots\nu_n}(p) &= \sum_{P(\mu)}\sum_{P(\nu)}\Biggl(\sum_{k=0}^{k_{\textrm{max},1}} \mathcal{A}_k^n\P_{\mu_1\mu_2}\P_{\nu_1\nu_2}\cdots\P_{\mu_{2k-1}\mu_{2k}}\P_{\nu_{2k-1}\nu_{2k}}\prod_{i=2k+1}^n\P_{\mu_i\nu_i} \nonumber\\
&\quad\,+ \feynP{\P}_{\mu_1}\feynP{\P}_{\nu_1}\sum_{k=0}^{k_{\textrm{max},2}} \mathcal{B}_k^n\P_{\mu_2\mu_3}\P_{\nu_2\nu_3}\cdots\P_{\mu_{2k}\mu_{2k+1}}\P_{\nu_{2k}\nu_{2k+1}}\prod_{i=2k+2}^n\P_{\mu_i\nu_i}\Biggr),\label{projection-general-2}
\end{align}
\end{widetext}
with
\begin{align}
k_{\textrm{max},1} &= \begin{cases}
                     \frac{n}{2} & n\; \textrm{even}, \\
		     \frac{n-1}{2} & n\; \textrm{odd},
                     \end{cases}
\nonumber\\
k_{\textrm{max},2} &= \begin{cases}
                     \frac{n-2}{2} & n\; \textrm{even}, \\
		     \frac{n-1}{2} & n\; \textrm{odd}.
                     \end{cases}
\end{align}
The coefficients $\mathcal{A}_k^n$ and $\mathcal{B}_k^n$ read
\begin{subequations}
\begin{align}
\mathcal{A}_k^n &= \frac{1}{(-2)^k}\frac{1}{n!k!(n-2k)!}\frac{(2n-2k+1)!!}{(2n+1)!!},\\
\mathcal{B}_k^n &= -\frac{1}{(-2)^k}\frac{1}{n!k!(n-2k-1)!}\frac{(2n-2k-1)!!}{(2n+1)!!}.
\end{align}
\end{subequations}
The spin-$\left(n+{1/2}\right)$ projection operator is totally symmetric in its $\mu_i$ indices as well as in its $\nu_j$ indices and satisfies the R-S constraints
\begin{align}
\gamma^{\mu_1}\P^{n+{1/2}}_{\mu_1\ldots\mu_n;\nu_1\ldots\nu_n}(p) &= \P^{n+{1/2}}_{\mu_1\ldots\mu_n;\nu_1\ldots\nu_n}(p)\gamma^{\nu_1} = 0, \nonumber\\
p^{\mu_1}\P^{n+{1/2}}_{\mu_1\ldots\mu_n;\nu_1\ldots\nu_n}(p) &= \P^{n+{1/2}}_{\mu_1\ldots\mu_n;\nu_1\ldots\nu_n}(p)p^{\nu_1} = 0.\label{P-RS_constraints}
\end{align}
The general expression for the spin-$\left(n+{1/2}\right)$ propagator, which was derived by Huang \ea in Ref.~\cite{Huang:2005js}, consists of a covariant part and a non-covariant part. The covariant part of the propagator is given as
\begin{align}
P_{\mu_1\ldots\mu_n;\nu_1\ldots\nu_n}(p) = \frac{\feynp{p} + m}{p^2 - m^2}\widetilde{\P}^{n+{1/2}}_{\mu_1\ldots\mu_n;\nu_1\ldots\nu_n}(p).
\end{align}
The ``on-shell'' spin projection operator $\widetilde{\P}^{n+{1/2}}_{\mu_1\ldots\mu_n;\nu_1\ldots\nu_n}(p)$ is obtained from the ``off-shell'' spin projection operator $\P^{n+{1/2}}_{\mu_1\ldots\mu_n;\nu_1\ldots\nu_n}(p)$ through the substitutions
\begin{subequations}
\label{projection_to_propagator}
\begin{align}
\P_{\mu\nu}(p) &= g_{\mu\nu} - \frac{1}{p^2}p_{\mu} p_{\nu}, \nonumber\\
&\rightarrow g_{\mu\nu} - \frac{1}{m^2}p_{\mu} p_{\nu},
\end{align}
and
\begin{align}
&\feynP{\P}_{\mu}(p)\feynP{\P}_{\nu}(p) \nonumber\\
&= \gamma_{\mu}\gamma_{\nu} + \frac{\feynp{p}}{p^2}\left(\gamma_{\mu} p_{\nu} - \gamma_{\nu} p_{\mu}\right) - \frac{1}{p^2}p_{\mu} p_{\nu}, \nonumber\\
&\rightarrow \gamma_{\mu}\gamma_{\nu} + \frac{1}{m}\left(\gamma_{\mu} p_{\nu} - \gamma_{\nu} p_{\mu}\right) - \frac{1}{m^2}p_{\mu} p_{\nu}.
\end{align}
\end{subequations}
These are equivalent to the substitutions \eq{P52_to_P} carried out for the spin-${5/2}$ theory. The non-covariant part of the propagator should obviously be ignored to preserve the Lorentz invariance of the transition amplitude, as is required by the special theory of relativity.

The consistency of gauge-invariant interactions, which are described by the interaction Lagrangian \eq{L_I-general}, can now be proven. As becomes clear from substitutions \eq{projection_to_propagator}, the spin-$\left(n+{1/2}\right)$ propagator $P_{\mu_1\ldots\mu_n;\nu_1\ldots\nu_n}(p)$ differs from the spin-$\left(n+{1/2}\right)$ projection operator $\P^{n+{1/2}}_{\mu_1\ldots\mu_n;\nu_1\ldots\nu_n}(p)$ in the momentum-dependent terms. However, considering the properties of Eq.~\eq{O-general-properties} the momentum-dependent terms of the propagator drop out from the interaction structure. As a consequence, the general interaction structure is invariant under substitutions of the type \eq{projection_to_propagator} and the propagator can be replaced by
\begin{align}
P_{\mu_1\ldots\mu_n;\nu_1\ldots\nu_n}(p) \rightarrow \frac{\feynp{p} + m}{p^2 - m^2}\P^{n+{1/2}}_{\mu_1\ldots\mu_n;\nu_1\ldots\nu_n}(p).
\end{align}
This proves the consistency of gauge-invariant interactions. The latter property follows directly from the fact that the physical spin-$\left(n+{1/2}\right)$ component $\P^{n+{1/2}}_{\mu_1\ldots\mu_n;\nu_1\ldots\nu_n}(\p)\psi^{\nu_1\ldots\nu_n}$ of $\psi_{\mu_1\ldots\mu_n}$ mediates the interaction. 

The expression for the most general consistent interaction structure is reduced to
\begin{widetext}
\begin{align}
&O^{n+{1/2}}_{(\mu_1\ldots\mu_n,\nu_1\ldots\nu_n)\sigma_1\ldots\sigma_n}(p)P^{\sigma_1\ldots\sigma_n;\tau_1\ldots\tau_n}(p)O^{n+{1/2}}_{(\lambda_1\ldots\lambda_n,\rho_1\ldots\rho_n)\tau_1\ldots\tau_n}(p) \nonumber\\
&= \frac{\feynp{p} + m}{p^2 - m^2}O^{n+{1/2}}_{(\mu_1\ldots\mu_n,\nu_1\ldots\nu_n)\sigma_1\ldots\sigma_n}(p)O^{n+{1/2}}_{(\lambda_1\ldots\lambda_n,\rho_1\ldots\rho_n)\tau_1\ldots\tau_n}(p) \nonumber\\
&\quad\,\times \sum_{P(\sigma)}\sum_{P(\tau)}\Biggl(\sum_{k=0}^{k_{\textrm{max},1}} \mathcal{A}_k^n g^{\sigma_1\sigma_2} g^{\tau_1\tau_2}\cdots g^{\sigma_{2k-1}\sigma_{2k}} g^{\tau_{2k-1}\tau_{2k}}\prod_{i=2k+1}^n g^{\sigma_i\tau_i} \nonumber\\
&\quad\,+ \gamma^{\sigma_1}\gamma^{\tau_1}\sum_{k=0}^{k_{\textrm{max},2}} \mathcal{B}_k^n g^{\sigma_2\sigma_3} g^{\tau_2\tau_3}\cdots g^{\sigma_{2k}\sigma_{2k+1}} g^{\tau_{2k}\tau_{2k+1}}\prod_{i=2k+2}^n g^{\sigma_i\tau_i}\Biggr),\label{interaction_structure-general}
\end{align}
\end{widetext}
owing to relations \eq{O-general-properties}. From Eq.~\eq{interaction_structure-general} it can be concluded that the power of the momentum dependence of the consistent interaction structure rises with the spin of the R-S field. This is a direct consequence of the gauge-invariance of the interaction. Indeed, the spin-specific momentum dependence of the consistent interaction structure is provided by the two interaction operators of Eq.~\eq{interaction_structure-general} and the interaction operator consists of products of $n$ four-momenta, as can be derived from the definition of $O^{n+{1/2}}_{(\mu_1\ldots\mu_n,\nu_1\ldots\nu_n)\sigma_1\ldots\sigma_n}(p)$ in Eq.~\eq{O-general}.

\subsubsection{Consistent couplings for the $(\phi\psi\psi^*_{\mu_1\ldots\mu_n})$- and $(A_\mu\psi\psi^*_{\mu_1\ldots\mu_n})$-theories}
\label{subsubsec:consistent_couplings}
Consistent interaction theories for off-shell spin-$\left(n+{1/2}\right)$ fields can be constructed from the gauge-invariant field $G_{\mu_1\ldots\mu_n,\nu_1\ldots\nu_n}$, which is defined as
\begin{align}
G_{\mu_1\ldots\mu_n,\nu_1\ldots\nu_n} &= \sum_{P(\mu)}\sum_{P(\nu)}\sum_{k=0}^{n}\mathcal{G}^n_k \p_{\nu_1}\cdots\p_{\nu_k} \nonumber\\
&\quad\,\times\p_{\mu_{k+1}}\cdots\p_{\mu_n}\psi_{\mu_1\ldots\mu_k\nu_{k+1}\ldots\nu_n}.
\end{align}
From $G_{\mu_1\ldots\mu_n,\nu_1\ldots\nu_n}$ a much more convenient gauge-invariant field can be derived, considering the fact that $G_{\mu_1\ldots\mu_n,\nu_1\ldots\nu_n}$ posesses twice as many Lorentz indices as the original field $\psi_{\mu_1\ldots\mu_n}$. This field is defined as
\begin{align}
\Psi_{\mu_1\ldots\mu_n} &= \gamma^{\nu_1}\cdots\gamma^{\nu_n}G_{\mu_1\ldots\mu_n,\nu_1\ldots\nu_n}, \nonumber\\
&= \sum_{P(\mu)}\sum_{k=0}^{n}n!\mathcal{G}^n_k \feynd{\p}^{\,k}\p_{\mu_{k+1}}\cdots\p_{\mu_n} \nonumber\\
&\quad\,\times\gamma^{\nu_{k+1}}\cdots\gamma^{\nu_n}\psi_{\mu_1\ldots\mu_k\nu_{k+1}\ldots\nu_n},\label{Psi-general}
\end{align}
and shares the properties of $\psi_{\mu_1\ldots\mu_n}$. Indeed, it has the same number of Lorentz indices, it is a totally symmetric field and it obeys the R-S constraints
\begin{subequations}
\label{Psi-RS_constraints}
\begin{align}
\p^{\mu_1}\Psi_{\mu_1\ldots\mu_n} &\rightarrow 0, \label{Psi-RS_constraints-1}\\
\gamma^{\mu_1}\Psi_{\mu_1\ldots\mu_n} &\rightarrow 0 \label{Psi-RS_constraints-2}.
\end{align}
\end{subequations}
That is the field combinations in \eq{Psi-RS_constraints} result in a zero transition amplitude, as will be clarified shortly. The corresponding interaction operator for the field $\Psi_{\mu_1\ldots\mu_n}$ reads
\begin{align}
&\O^{n+{1/2}}_{(\mu_1\ldots\mu_n)\lambda_1\ldots\lambda_n}(\p) \nonumber\\
&= \gamma^{\nu_1}\cdots\gamma^{\nu_n}O^{n+{1/2}}_{(\mu_1\ldots\mu_n,\nu_1\ldots\nu_n)\lambda_1\ldots\lambda_n}(\p), \nonumber\\
&= \frac{1}{n!}\sum_{P(\lambda)}\O^{{3/2}}_{(\mu_1)\lambda_1}(\p)  \cdots  \O^{{3/2}}_{(\mu_n)\lambda_n}(\p),
\end{align}
with
\begin{align}
\O^{{3/2}}_{(\mu)\lambda}(\p) &= \gamma^{\nu}\O^{{3/2}}_{(\mu,\nu)\lambda}(\p), \nonumber\\
&= i\left(\p_\mu\gamma_\lambda - \feynd{\p}g_{\mu\lambda}\right).
\end{align}
The advantage of using $\Psi_{\mu_1\ldots\mu_n}$ instead of $G_{\mu_1\ldots\mu_n,\nu_1\ldots\nu_n}$ should be clear: the reduction in the number of Lorentz indices along with the R-S constraints \eq{Psi-RS_constraints}, lower the number of possible interaction theories. The interaction structure associated with $\Psi_{\mu_1\ldots\mu_n}$ is found to be
\begin{align}
&\O^{n+{1/2}}_{(\mu_1\ldots\mu_n)\lambda_1\ldots\lambda_n}(p)P^{\lambda_1\ldots\lambda_n;\rho_1\ldots\rho_n}(p)\O^{n+{1/2}}_{(\nu_1\ldots\nu_n)\rho_1\ldots\rho_n}(p) \nonumber\\
&= p^{2n}\frac{\feynp{p} + m}{p^2 - m^2}\P^{n+{1/2}}_{\mu_1\ldots\mu_n;\nu_1\ldots\nu_n}(p). \label{interaction_structure-Psi}
\end{align}
Indeed, only the term $(-\feynp{p})^{n}g_{\mu_1\lambda_1}\cdots g_{\mu_n\lambda_n}$ of the interation operator, which corresponds to the term $(-i\feynd{\p})^{n}\psi_{\mu_1\ldots\mu_n}$ of $\Psi_{\mu_1\ldots\mu_n}$, contributes to the interaction structure. All other terms of the interaction operator contain at least one $\gamma_{\lambda_i}$ and vanish subsequently due to the first of properties \eq{P-RS_constraints}. Since the interaction structure associated with $\Psi_{\mu_1\ldots\mu_n}$ is proportional to the spin-$\left(n+{1/2}\right)$ projection operator $\P^{n+{1/2}}_{\mu_1\ldots\mu_n;\nu_1\ldots\nu_n}(p)$, relations \eq{Psi-RS_constraints} follow immediately from properties \eq{P-RS_constraints}.
Eq.~\eq{interaction_structure-Psi} is just how a consistent local interaction structure would be constructed in an intuitive way. It is proportional to the spin-$\left(n+{1/2}\right)$ projection operator, which ensures the consistency of the interaction, and the non-localities of the latter are exactly canceled through the $p^{2n}$ factor. 

In order to construct consistent couplings for the $(\phi\psi\psi^*_{\mu_1\ldots\mu_n})$- and $(A_\mu\psi\psi^*_{\mu_1\ldots\mu_n})$-theories, the spin-${3/2}$ theory is considered first, since this is the most simple and hence most studied R-S theory. Popular choices for the $(\phi\psi\psi^*_\mu)$- and $(A_\mu\psi\psi^*_\mu)$-couplings read \cite{Janssen:2001wk}
\begin{align}
\L'_{\phi\psi\psi^*_\mu} = \frac{ig_0}{m_\phi}\overline{\psi}^\mu\Theta_{\mu\nu}(z_0)\Gamma\psi\p^\nu\phi + \textrm{H.c.},\label{32-scalar}
\end{align}
and
\begin{align}
{\L'}_{A_\mu\psi\psi^*_\mu}^{(1)} &= \frac{ig_1}{2m_\psi}\overline{\psi}^\mu\Theta_{\mu\nu}(z_1)\Gamma \gamma_\lambda \psi F^{\lambda\nu} + \textrm{H.c.}, \nonumber\\
{\L'}_{A_\mu\psi\psi^*_\mu}^{(2)} &= -\frac{g_2}{4m_\psi^2}\overline{\psi}^{\mu}\Theta_{\mu\nu}(z_2)\Gamma \p_\lambda \psi F^{\lambda\nu} + \textrm{H.c.}, \nonumber\\
{\L'}_{A_\mu\psi\psi^*_\mu}^{(3)} &= -\frac{g_3}{4m_\psi^2}\overline{\psi}^{\mu}\Theta_{\mu\nu}(z_3)\Gamma \psi \p_\lambda F^{\lambda\nu} + \textrm{H.c.}.\label{32-vector}
\end{align}
Here, the $g_i$'s are coupling constants, the $z_i$'s are so-called off-shell parameters, $\Theta_{\mu\nu}(z) = g_{\mu\nu} - \left(z+\frac{1}{2}\right)\gamma_\mu\gamma_\nu$ is the off-shell tensor and $F_{\mu\nu} = \p_\mu A_\nu - \p_\nu A_\mu$. Further, $\Gamma = 1$ if the parity of the outgoing state and the incoming state are equal, and $\Gamma = \gamma_5$ in the opposite situation. The Lagrangians of Eqs.~\eq{32-scalar} and \eq{32-vector} are inconsistent since they involve unphysical interactions which are mediated by the spin-${1/2}$ component of $\psi_\mu$. This comes as no surprise since these couplings are not gauge-invariant. However, the inconsistent off-shell interactions can be turned into consistent interactions by means of the substitution
\begin{align}
\Theta_{\mu\nu}(z_i)\psi^\nu \rightarrow \frac{1}{m}\Psi_\mu,
\end{align}
where $m$ represents the mass of $\phi$ or twice the mass of $\psi$. The new, consistent $(\phi\psi\psi^*_\mu)$- and $(A_\mu\psi\psi^*_\mu)$-couplings read
\begin{align}
\L_{\phi\psi\psi^*_\mu} = \frac{ig_0}{m_\phi^2}\overline{\Psi}_\mu\Gamma\psi\p^\mu\phi + \textrm{H.c.},\label{consistent-scalar}
\end{align}
and
\begin{align}
\L_{A_\mu\psi\psi^*_\mu}^{(1)} &= \frac{ig_1}{4m_\psi^2}\overline{\Psi}_\mu\Gamma \gamma_\nu \psi F^{\nu\mu} + \textrm{H.c.}, \nonumber\\
\L_{A_\mu\psi\psi^*_\mu}^{(2)} &= -\frac{g_2}{8m_\psi^3}\overline{\Psi}_\mu\Gamma \p_\nu \psi F^{\nu\mu} + \textrm{H.c.}, \nonumber\\
\L_{A_\mu\psi\psi^*_\mu}^{(3)} &= -\frac{g_3}{8m_\psi^3}\overline{\Psi}_\mu\Gamma \psi \p_\nu F^{\nu\mu} + \textrm{H.c.}.\label{consistent-vector}
\end{align}
The consistency of these interactions is guaranteed by the explicitly gauge-invariant field $\Psi_\mu = i\left(\p_\mu\gamma^\nu\psi_\nu - \feynd{\p}\psi_\mu\right)$. Note that the derivative acting on $\phi$ in Lagrangian \eq{consistent-scalar} cannot be replaced by a Dirac matrix due to property \eq{Psi-RS_constraints-1}. The interaction Lagrangian with the derivative acting on $\psi$, i.e.\
\begin{align}
\L_{\phi\psi\psi^*_\mu} = \frac{ig_0}{m_\phi^2}\overline{\Psi}_\mu\Gamma\p^\mu\psi\phi + \textrm{H.c.},
\end{align}
is equivalent to \eq{consistent-scalar}, aside from a minus sign, as is seen from partial integration and property \eq{Psi-RS_constraints-2}.

The consistent interaction Lagrangians for the spin-${3/2}$ theory, i.e.~\eq{consistent-scalar} and \eq{consistent-vector}, can be generalized to the spin-$\left(n+{1/2}\right)$ theory. The consistent $(\phi\psi\psi^*_{\mu_1\ldots\mu_n})$- and $(A_\mu\psi\psi^*_{\mu_1\ldots\mu_n})$-couplings read
\begin{widetext}
\begin{align}
\L_{\phi\psi\psi^*_{\mu_1\ldots\mu_n}} = \frac{i^ng_0}{m_\phi^{2n}}\overline{\Psi}_{\mu_1\ldots\mu_n}\Gamma\psi\p^{\mu_1}\cdots\p^{\mu_n}\phi + \textrm{H.c.},\label{consistent-scalar-general}
\end{align}
and
\begin{align}
\L_{A_\mu\psi\psi^*_{\mu_1\ldots\mu_n}}^{(1)} &= \frac{i^ng_1}{(2m_\psi)^{2n}}\overline{\Psi}_{\mu_1\ldots\mu_{n-1}\mu_n}\Gamma \gamma_\nu\p^{\mu_1}\cdots\p^{\mu_{n-1}} \psi F^{\nu\mu_n} + \textrm{H.c.}, \nonumber\\
\L_{A_\mu\psi\psi^*_{\mu_1\ldots\mu_n}}^{(2)} &= \frac{i^{n+1}g_2}{(2m_\psi)^{2n+1}}\overline{\Psi}_{\mu_1\ldots\mu_{n-1}\mu_n}\Gamma \p_\nu \p^{\mu_1}\cdots\p^{\mu_{n-1}}\psi F^{\nu\mu_n} + \textrm{H.c.}, \nonumber\\
\L_{A_\mu\psi\psi^*_{\mu_1\ldots\mu_n}}^{(3)} &= \frac{i^{n+1}g_3}{(2m_\psi)^{2n+1}}\overline{\Psi}_{\mu_1\ldots\mu_{n-1}\mu_n}\Gamma \p^{\mu_1}\cdots\p^{\mu_{n-1}}\psi \p_\nu F^{\nu\mu_n} + \textrm{H.c.}. \label{consistent-vector-general}
\end{align}
\end{widetext}
Indeed, the joining $(n-1)$ Lorentz indices of $\Psi_{\mu_1\ldots\mu_n}$, as compared to $\Psi_\mu$, can only be contracted by derivatives which do not act on $\Psi_{\mu_1\ldots\mu_n}$ since properties \eq{Psi-RS_constraints} hold.

\section{Consistent interactions in hadron physics}
\label{sec:hadron_physics}
The nucleon has various excited states, which are commonly dubbed ``nucleon resonances''. They are identified with their different masses, spins and decay widths, which reflect the specific internal structure of the resonance. The quantitative information on the excited nucleon states is gathered by PDG from various partial-wave analyses, which aim at describing pion- and photon-induced meson production processes \cite{Nakamura:2010zzi}. Alternatively, these processes can be described by isobar models. An extensive overview of the various partial-wave and isobar models is found in Ref.~\cite{Klempt:2009pi}. In order to describe nucleon excitation processes in a consistent way, isobar models require a consistent, high-spin interaction theory. Such a theory was developed in Sec.~\ref{subsubsec:consistent_couplings}. In the next sections it is illustrated how this interaction theory can be implemented in the study of processes which are of key importance in hadron physics. To this end, the $K^+\Lambda$ photoproduction process from the proton is selected
\begin{align}
\gamma p \rightarrow K^+\Lambda.
\end{align}
It is worth stressing that all discussions of the next sections apply equally well to other hadronic processes that involve off-shell high-spin interactions.

\subsection{Inconsistency of standard hadronic form factors}
\label{subsec:standard_HFF}
The threshold energy $W_0$ for $K^+\Lambda$ production is given by \cite{Nakamura:2010zzi}
\begin{align}
W_0 = m_{K^+} + m_{\Lambda} \approx 1610 \; \textrm{MeV}.
\end{align}
In an effective-field framework, the $p(\gamma,K^+)\Lambda$ reaction is modeled with hadrons as the basic degrees of freedom, i.e\ the hadrons are represented by effective quantum fields. The expression for the differential $p(\gamma,K^+)\Lambda$ cross section in the center-of-mass frame is given by \cite{Janssen:2001pe}
\begin{align}
\frac{\d\sigma}{\d\Omega_K} = \frac{1}{64\pi^2}\frac{1}{W^2}\frac{|\vec{p}_K|}{E_\gamma^{\textrm{lab}}}\overline{\sum}_{\lambda, \lambda_p, \lambda_\Lambda}\bigl|\mathcal{M}_{\lambda,\lambda_p,\lambda_\Lambda} \bigr|^2. \label{diffcs}
\end{align}
Here, $W$ is the invariant mass, $E_\gamma^{\textrm{lab}}$ is the photon energy in the laboratory frame, $\vec{p}_K$ is the three-momentum and $\theta_K$ the scattering angle of the kaon in the center-of-mass frame. Further, $\lambda$, $\lambda_p$ and $\lambda_\Lambda$ are the photon, proton and hyperon polarizations. The notation $\overline{\sum}_{\lambda, \lambda_p, \lambda_\Lambda}$ denotes appropriate summing and/or averaging over the polarizations of initial- and final-state particles. Finally, $\mathcal{M}_{\lambda,\lambda_p,\lambda_\Lambda}$ is the total transition amplitude. Its squared absolute value is calculated as
\begin{align}
\bigl|\mathcal{M}_{\lambda,\lambda_p,\lambda_\Lambda} \bigr|^2 = \bigl|\overline{u}_\Lambda^{\lambda_\Lambda}T^{\mu}\varepsilon_\mu^\lambda u_p^{\lambda_p}\bigr|^2,
\end{align}
with $u_p^{\lambda_p}$ and $u_\Lambda^{\lambda_\Lambda}$ the proton and hyperon spinors, $\varepsilon_\mu^\lambda$ the polarization four-vector of the photon and $T_\mu$ the truncated interaction current.

\begin{figure}[t]
\centering
\includegraphics[scale=.9,bb=200 492 395 350,page=1,clip=true]{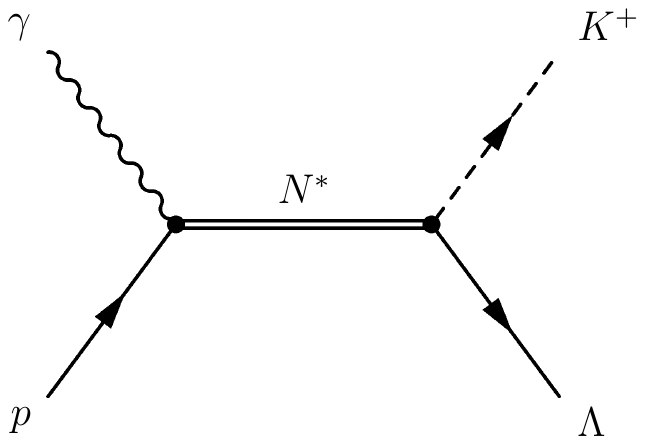}
\vspace{120pt}
\caption{The tree-level Feynman diagram for the resonant $s$ channel contribution to the $p(\gamma,K^+)\Lambda$ reaction in an effective-field theoretical framework.}\label{tree-level}
\end{figure}

Fig.~\ref{tree-level} depicts the tree-level Feynman diagram for the resonant $s$ channel contribution to the $p(\gamma,K^+)\Lambda$ reaction.
In this channel, the photon excites the proton to form a nucleon resonance, which is labeled as $N^*$. In the $N^*$'s rest frame the spin of the $N^*$ can be determined by the photon-proton relative orbital angular momentum. The nucleon resonance subsequently decays into a $K^+$ and a $\Lambda$. In order to account for the finite lifetime of the $N^*$, the following substitution in the expression for the $N^*$ propagator is required
\begin{align}
\frac{\feynp{p}_R + m_R}{p_R^2 - m_R^2} \rightarrow \frac{\feynp{p}_R + m_R}{p_R^2 - m_R^2+im_R\Gamma_R}.\label{decay_width}
\end{align}
In the tree-level approximation of Fig.~\ref{tree-level}, the propagator remains ``undressed''. As a consequence, the decay width of the unstable particle, i.e.\ the resonance, is not generated dynamically. The finite lifetime of the resonance can then be implemented by means of the substitution of Eq.~\eq{decay_width}.

\begin{figure}[t]
\centering
\includegraphics[scale=0.4]{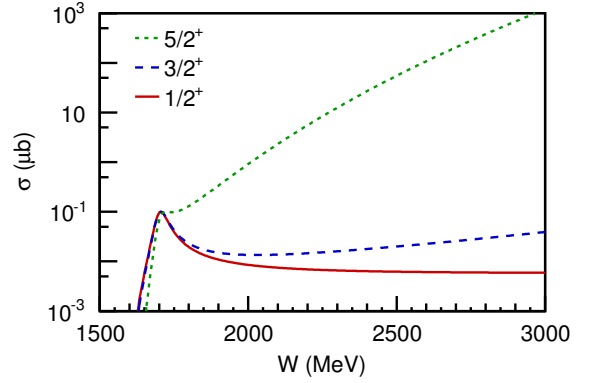}
\caption{The energy dependence of the $\gamma p \rightarrow N^* \rightarrow K^+\Lambda$ cross section. The $N^*$ is a mock resonance with $m_R = 1700$ MeV, $\Gamma_R = 50$ MeV and $J_R^P = 1/2^+$ (solid curve), $3/2^+$ (dashed curve), $5/2^+$ (dotted curve). The coupling constants are $g_{1/2} =1.7$, $g_{3/2} = 0.75$ and $g_{5/2} = 3.2$.} \label{spin_12-32-52}
\end{figure}

From expression \eq{diffcs}, the unpolarized total cross section
\begin{align}
\sigma(W) = \int\frac{\d\Omega_K}{64\pi^2}\frac{1}{W^2}\frac{|\vec{p}_K|}{E_\gamma^{\textrm{lab}}}\frac{1}{2^2}\sum_{\lambda, \lambda_p, \lambda_\Lambda}\bigl|\mathcal{M}_{\lambda,\lambda_p,\lambda_\Lambda} \bigr|^2,\label{tds}
\end{align}
can be calculated. In the presented calculations the coupling constants of Eqs.~\eq{consistent-scalar-general} and \eq{consistent-vector-general} are chosen to be equal, i.e.\ $g_0 = g_1 = g_2 = g$. The coupling constant $g$ is determined from the requirement that $\sigma_{\max} = 0.10$ $\mu$b, which is a typical order of magnitude for the reaction $\gamma p \rightarrow K^+\Lambda$. In Fig.~\ref{spin_12-32-52}, $\sigma$ is plotted for three artificial resonances with $m_R = 1700$ MeV, $\Gamma_R = 50$ MeV and spins $J_R^P = 1/2^+$, $3/2^+$, $5/2^+$. The coupling constant $g$ for each of the three resonances is denoted by $g_{1/2}$, $g_{3/2}$ and $g_{5/2}$.
The explicit expressions for the spin-$1/2^+$, spin-$3/2^+$ and spin-$5/2^+$ truncated currents are given by
\begin{align}
T^{1/2^+}_\mu &= (g_0\gamma_5)\biggl(\frac{\feynp{p}_R + m_R}{p_R^2 - m_R^2+im_R\Gamma_R}\biggr) \nonumber\\
&\quad\,\times\biggl(\frac{ieg_1}{2m_p}\Bigl(k_\mu -\feynk{k}\gamma_\mu\Bigr)\biggr),\label{truncated_current-12}
\end{align}
\begin{align}
T^{3/2^+}_\mu &= \biggl(\frac{ig_0}{m_{K^+}^2}p_{K^+}^{\mu_1}\biggr) \nonumber\\
&\quad\,\times\biggl(p_R^2\frac{\feynp{p}_R + m_R}{p_R^2 - m_R^2+im_R\Gamma_R}\P^{{3/2}}_{\mu_1;\nu_1}(p_R)\biggr)\nonumber\\
&\quad\,\times\biggl(\frac{ieg_1}{4m_p^2}\gamma_5\Bigl(\feynk{k}g^{\nu_1}_\mu-k^{\nu_1}\gamma_\mu\Bigr) \nonumber\\
&\quad\,+ \frac{ieg_2}{8m_p^3}\gamma_5\Bigl((k\cdot p)g^{\nu_1}_\mu-k^{\nu_1}p_\mu\Bigr)\biggr),\label{truncated_current-32}
\end{align}
\begin{align}
T^{5/2^+}_\mu &= \biggl(\frac{ig_0}{m_{K^+}^4}\gamma_5p_{K^+}^{\mu_1}p_{K^+}^{\mu_2}\biggr) \nonumber\\
&\quad\,\times\biggl(p_R^4\frac{\feynp{p}_R + m_R}{p_R^2 - m_R^2+im_R\Gamma_R}\P^{{5/2}}_{\mu_1\mu_2;\nu_1\nu_2}(p_R)\biggr) \nonumber\\
&\quad\,\times\biggl(\frac{ieg_1}{16m_p^4}p^{\nu_1}\Bigl(\feynk{k}g^{\nu_2}_\mu-k^{\nu_2}\gamma_\mu\Bigr) \nonumber\\
&\quad\,+ \frac{ieg_2}{32m_p^5}p^{\nu_1}\Bigl((k\cdot p)g^{\nu_2}_\mu-k^{\nu_2}p_\mu\Bigr)\biggr),\label{truncated_current-52}
\end{align}
and the corresponding cross sections are denoted as $\sigma_{1/2^+}$, $\sigma_{3/2^+}$ and $\sigma_{5/2^+}$. In the above expressions, $e$ represents the charge of the proton, $k_\mu$, $p_\mu$, $p_{K^+,\mu}$ and $p_{R,\mu} = k_\mu + p_\mu$ the four-momenta of the photon, the proton, the kaon, and the $N^*$. Further, $m_p$ and $m_{K^+}$ are the masses of the proton and the kaon. Note that $p_R^2 = W^2$. The expression for $T^{1/2^+}_\mu$ was obtained from Ref.~\cite{Benmerrouche:1994uc}. The expressions for $T^{3/2^+}_\mu$ and $T^{5/2^+}_\mu$ were derived from the consistent interaction Lagrangians \eq{consistent-scalar-general} and \eq{consistent-vector-general}, and the consistent interaction structure \eq{interaction_structure-Psi}. Since the equations of motion for the real photon field are given by
\begin{align}
\p_\mu F^{\mu\nu} = 0,
\end{align}
the truncated currents $T^{3/2^+}_\mu$ and $T^{5/2^+}_\mu$ do not receive contributions from the third Lagrangian of Eq.~\eq{consistent-vector-general}. 

Inspecting Fig.~\ref{spin_12-32-52} it is observed that $\sigma_{1/2^+}$ decreases asymptotically with $W$ whereas $\sigma_{3/2^+}$ and $\sigma_{5/2^+}$ grow indefinitely with $W$. The higher the spin of the resonance the faster $\sigma$ rises with $W$. This clearly is an unphysical and an unacceptable behavior. In order to remedy this one commonly introduces a hadronic form factor, which cuts the transition amplitude $\mathcal{M}_{\lambda,\lambda_p,\lambda_\Lambda}$ beyond a certain energy scale. Hadronic form factors which are commonly used in the literature are of the dipole form \cite{Pearce:1990uj}
\begin{align}
F_d(s;m_R,\Lambda_R) = \frac{\Lambda_R^4}{(s-m_R^2)^2+\Lambda_R^4},\label{d_HFF}
\end{align}
or the Gaussian form \cite{Corthals:2005ce}
\begin{align}
F_G(s;m_R,\Lambda_R) = \exp\left(-\frac{(s-m_R^2)^2}{\Lambda_R^4}\right),\label{G_HFF}
\end{align}
with $s = W^2$ and $\Lambda_R$ the cut-off energy.

\begin{figure}[t]
\centering
\subfigure{\includegraphics[scale=0.4]{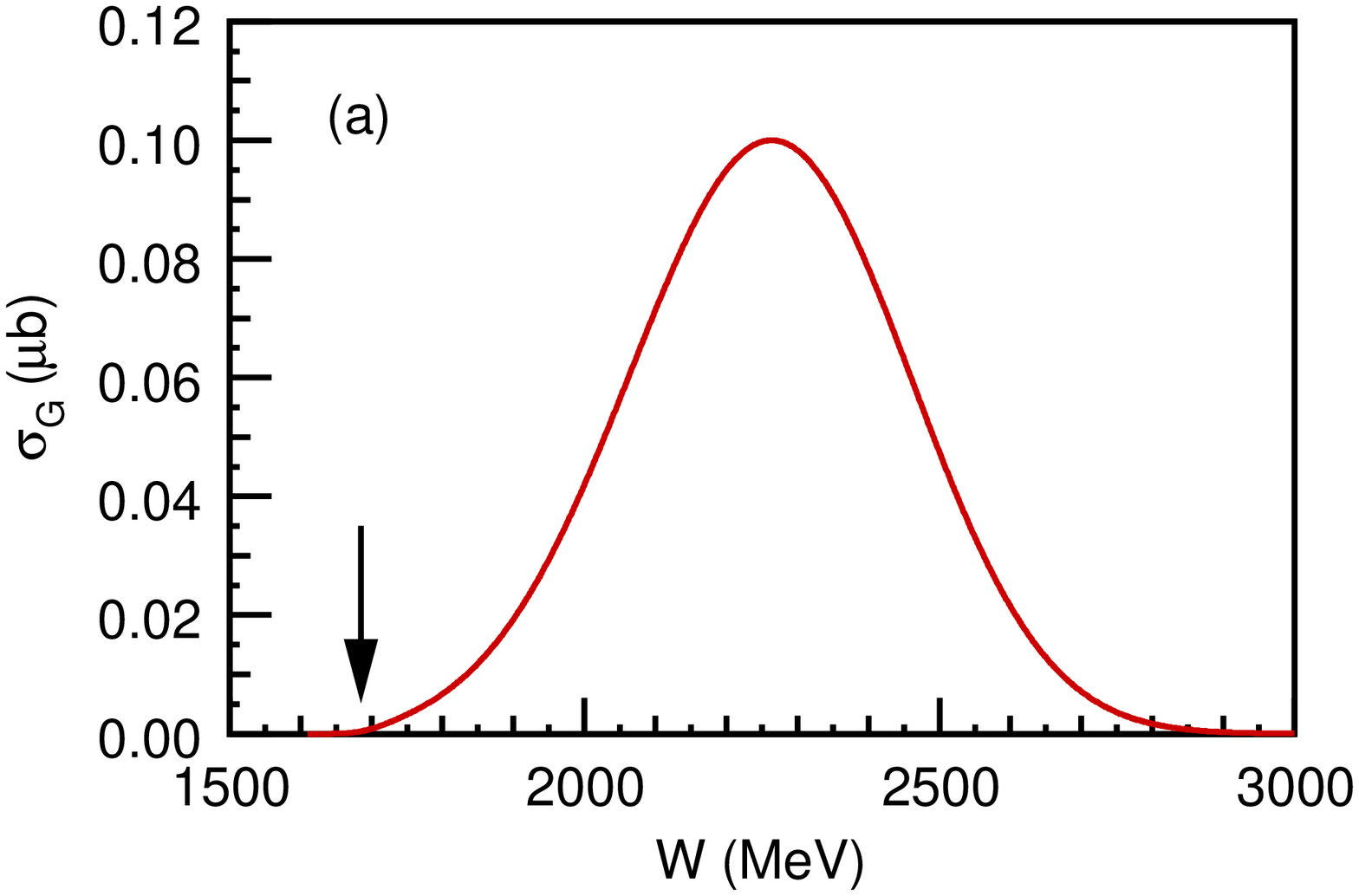}}
\subfigure{\includegraphics[scale=0.4]{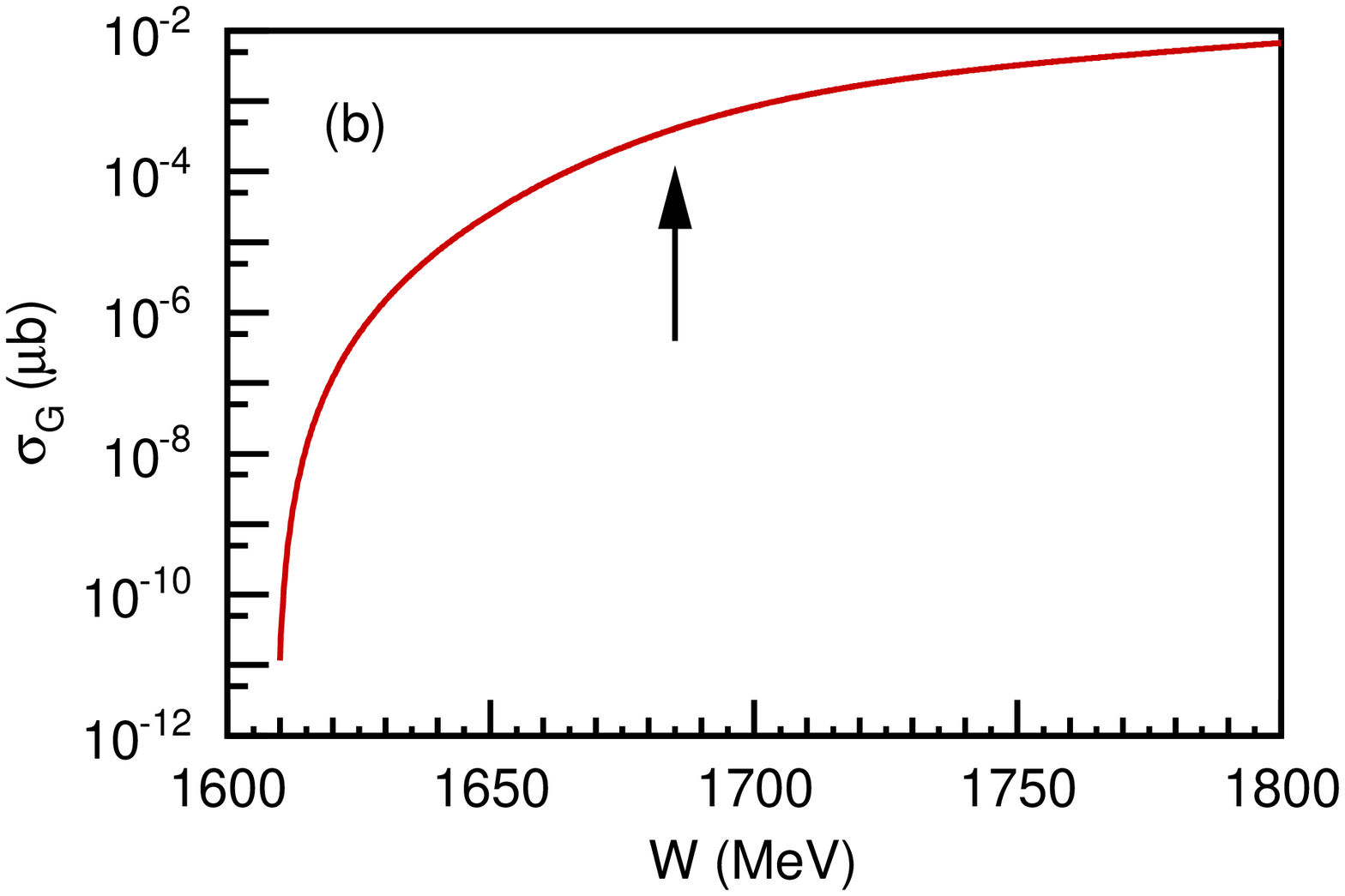}}
\caption{(a) The energy dependence of the $\gamma p \rightarrow N(1680)\;F_{15} \rightarrow K^+\Lambda$ cross section. A Gaussian form factor was used with $\Lambda_R = 1500$ MeV and $g = 1.7$. (b) A semi-logarithmic view of the cross section in the threshold-energy region. The arrows indicate the position of the mass of the \mbox{$N(1680)\;F_{15}$}.} \label{spin-52_bumps}
\end{figure}

For the remainder of this discussion, the following resonant $s$-channel contribution to the $p(\gamma,K^+)\Lambda$ reaction will be investigated in detail
\begin{align}
\gamma p \rightarrow N(1680)\;F_{15} \rightarrow K^+\Lambda.
\end{align}
The nucleon resonance \mbox{$N(1680)\;F_{15}$} is an established $J^P_R = {5/2}^+$ resonance with a four-star rating in the Review of Particle Physics of PDG \cite{Nakamura:2010zzi}. It has a mass $m_R = 1685$ MeV and a decay width $\Gamma_R = 130$ MeV. The computed cross section that uses a Gaussian form factor to cut off the transition amplitude at high energies, is denoted as $\sigma_G$. The value of the cut-off energy is fixed at a typical value of $\Lambda_R = 1500$ MeV. The result of the calculation of $\sigma_G$ is shown in Fig.~\ref{spin-52_bumps}(a). 
A seemingly resonant structure is observed. This structure, however, is not associated with the resonant structure of the \mbox{$N(1680)\;F_{15}$}. Indeed, $\sigma_G$ has $W_{\max} \approx 2250$ MeV and $\textsc{fwhm} \approx 450$ MeV, with $W_{\max}$ the invariant mass that corresponds to the maximum value of the cross section and \textsc{fwhm} the full width at half maximum. Clearly, the computed energy dependence of $\sigma_G$ displays little resemblance to the expected behavior of a resonance with $m_R = 1685$ MeV and $\Gamma_R = 130$ MeV. Fig.~\ref{spin-52_bumps}(b), which provides a closer look at $\sigma_G$ in the threshold-energy region, reveals that any sign of a resonant structure at $W \approx m_R$ is missing.

The observed energy dependence of $\sigma_G$, which can be conceived as unphysical, is generated by the combination of the opposite high-energy behavior of $\sigma$ and $F_G$. In the high-energy limit, $\sigma$ rises with the energy as is observed in Fig.~\ref{spin_12-32-52}. This feature is characteristic for off-shell high-spin interactions. On the other hand, $F_G$ decreases for growing $W > m_R$. For a particular value of the invariant mass, denoted by $W_{\max}$, the decrease of $F_G$ becomes strong enough so as to prevent $\sigma_G$ from growing indefinitely. As a result, the maximum value of $\sigma_G$ is reached at $W_{\max}$ and an artificial structure is created. The fact that the resonant structure at $W \approx m_R$ is not observed in the computed energy dependence of $\sigma_G$, can be attributed to the relatively large decay width of the \mbox{$N(1680)\;F_{15}$} in combination with the fast increase of $\sigma$ with growing $W$.

\begin{figure}[t]
\centering
\subfigure{\includegraphics[scale=0.4]{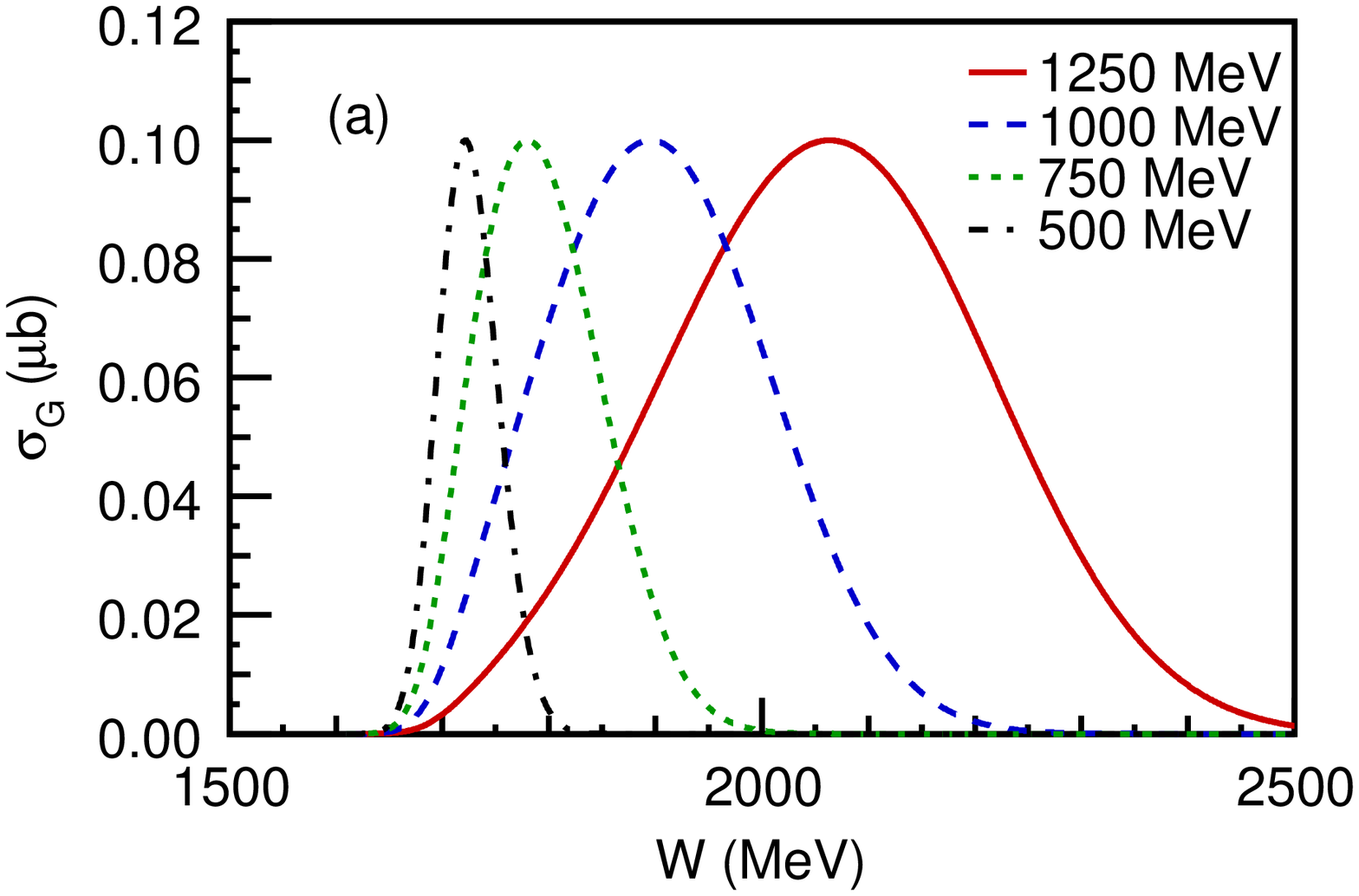}}
\subfigure{\includegraphics[scale=0.4]{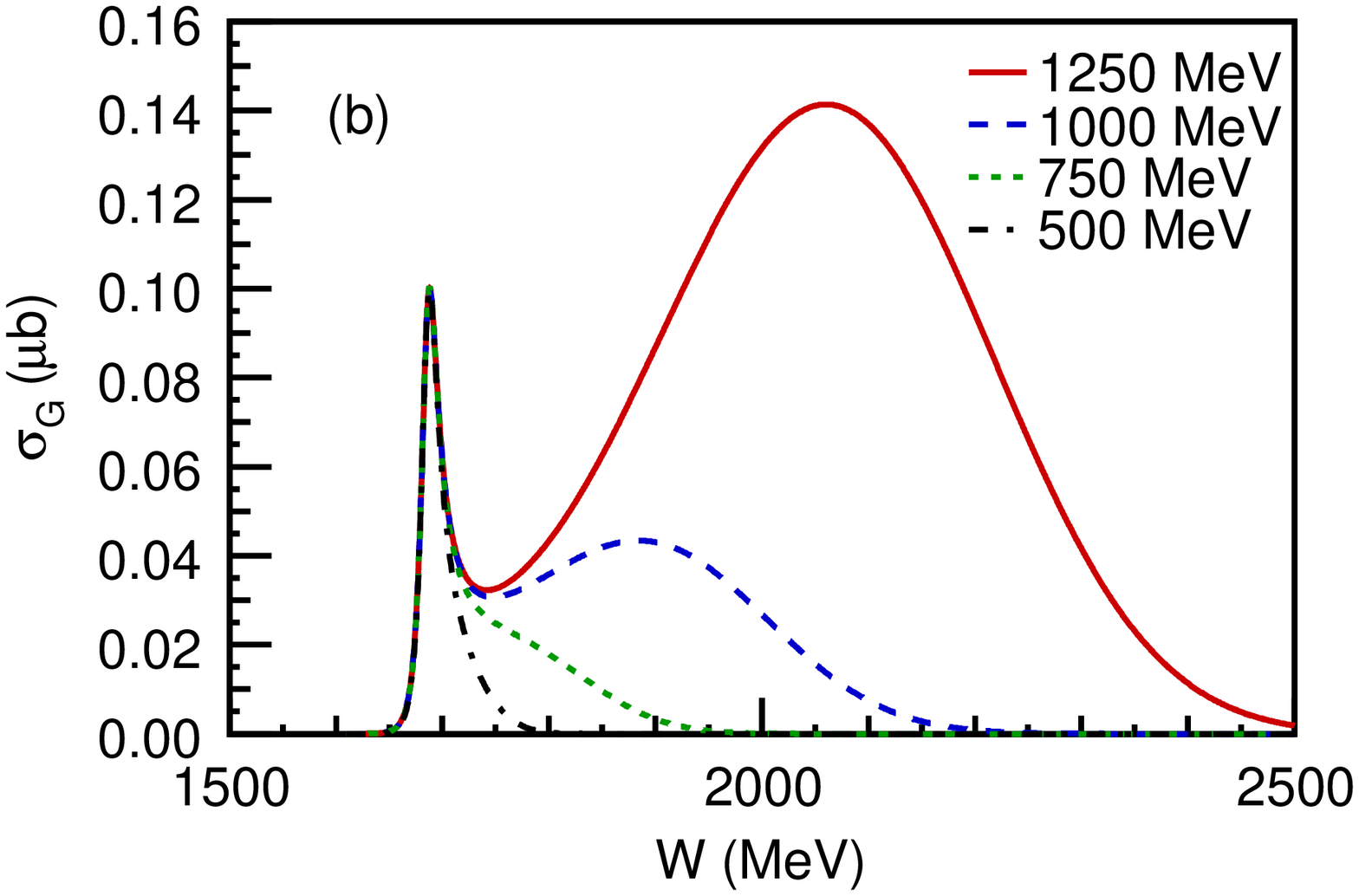}}
\caption{The energy dependence of the $\gamma p \rightarrow N(1680)\;F_{15} \rightarrow K^+\Lambda$ cross section for various values of the cut-off energy of the Gaussian form factor. In (a) the real decay width of the \mbox{$N(1680)\;F_{15}$} was used, i.e.\ $\Gamma_R = 130$ MeV, and $g =$ 2.4, 3.3, 4.2, 5.3 for $\Lambda_R =$ 1250, 1000, 750, 500 MeV. In (b) the decay width of the \mbox{$N(1680)\;F_{15}$} was set to $\Gamma_R' = 20$ MeV and $g =$ 2.6.} \label{spin-52_bumps-shift}
\end{figure}

Clearly, the computed cross section $\sigma_G$ of Fig.~\ref{spin-52_bumps}(a) lacks any obvious physical meaning. Indeed, the resonant structure at $W \approx m_R$ is erased and an unphysical bump dominates $\sigma_G$ for $m_R$ $\lesssim W \lesssim$ $3000$ MeV. A clear-cut remedy consists of cutting off the physics at smaller energies, i.e.\ lowering $\Lambda_R$. Fig.~\ref{spin-52_bumps-shift}(a) illustrates the energy dependence of $\sigma_G$ for a range of cut-off energies. 
Indeed, it is observed that lowering $\Lambda_R$ results in a mere shift of the unphysical bump towards the threshold energy $W_0$. The peak position and width of the bump in the energy dependence of $\sigma_G$ appear to be a function of $\Lambda_R$. The unphysical bump persistently dominates $\sigma_G$. In Fig.~\ref{spin-52_bumps-shift}(b) the decay width of the \mbox{$N(1680)\;F_{15}$} was artificially lowered to $\Gamma_R' = 20$ MeV so that the \mbox{$N(1680)\;F_{15}$}'s resonance peak is dominant at $W \approx m_R$. In this case lowering $\Lambda_R$ is indeed an effective remedy. However, it is not a priori clear what value that should be assigned to $\Lambda_R$. Furthermore, most of the established nucleon resonances, if not all, have a relatively large decay width \cite{Nakamura:2010zzi}. As such, lowering $\Lambda_R$ does not really provide a physically acceptable solution. Similar problems occur for spin-${3/2}$ resonances and higher-spin resonances. This is a feature that is inherent to the consistent description of high-spin interactions within the R-S framework. The conclusion is that the unphysical structure in $\sigma_G$ cannot be removed in a consistent way by lowering the value of the form factor's cut-off energy.

It is important to stress that with the dipole form factor of Eq.~\eq{d_HFF} things get even worse. Indeed, the dipole form factor does not fall fast enough with energy and $\sigma_d$ keeps on growing in the high-energy limit. This is obviously an unsatisfactory situation.

In Ref.~\cite{Shklyar:2009cx}, consistent spin-${5/2}$ interactions are constructed from the gauge-invariant field $\Psi'_{\mu\nu}$, defined as
\begin{align}
\Psi'_{\mu\nu} = \frac{\dalembert^2}{m^4}\P^{{5/2}}_{\mu\nu;\lambda\rho}(\p)\psi^{\lambda\rho},
\end{align}
with $m$ an arbitrary mass. The gauge-invariant field $\Psi'_{\mu\nu}$ is restricted to a pure spin-${5/2}$ component. This restriction is indispensable since the spin-${5/2}$ propagator of Ref.~\cite{Shklyar:2009cx} does not generate consistent couplings from interaction theories that are only required to be invariant under the uRS$_{5/2}$ \eq{RS-gauge-2}, as is clarified in Sec.~\ref{subsubsec:propagator-interaction_structure}. The gauge-invariant field $\Psi'_{\mu\nu}$ gives rise to the following truncated current for the tree-level $p(\gamma,K^+)\Lambda$ amplitude
\begin{align}
T'_\mu &= \biggl(-\frac{p_R^2}{m_{K^+}^2}\biggr)T_\mu\biggl(-\frac{p_R^2}{4m_p^2}\biggr), \nonumber\\
&= \frac{W^4}{4m_p^2m_{K^+}^2}T_\mu,\label{truncated_current-2}
\end{align}
with $T_\mu$ defined in Eq.~\eq{truncated_current-52}. Since it features additional powers of $W$, the corresponding cross section grows even stronger with $W$ in the high-energy limit than applying the consistent interaction theory of Sec.~\ref{subsubsec:consistent_couplings}. It is clear that its unphysical behavior is even more problematic.

\subsection{The multidipole-Gauss form factor}
\label{subsec:multidipole-Gauss_form_factor}
In Sec.~\ref{subsec:standard_HFF} it was pointed out that the unphysical behavior of $\sigma_G$ is caused by the divergent high-energy behavior of $\sigma$ in the tree-level approximation. This feature is characteristic for consistent high-spin interactions. In order to restore the physical resonance peak, the high-energy behavior of $\sigma$ needs to be regulated. The angular dependence of $\sigma$, which reflects the quantum numbers of the exchanged particles, should be left unaltered by such an operation. 

In the expression for the consistent interaction structure \eq{interaction_structure-Psi}, the prefactor $(p_R^2)^{n_R} = s^{n_R}$, with $n_R = J_R-\frac{1}{2}$, combined with the $2n_R$ four-momenta that are to be contracted with the spin-$J_R$ projection operator, give rise to a factor $s^{2n_R} = (s^2)^{J_R-\frac{1}{2}}$ for $s \gg m_R^2$. It is exactly this factor that causes the unphysical behavior of $\sigma_G$. In order to resolve these problems, the $(s^2)^{J_R-\frac{1}{2}}$ factor needs to be included in the denominator of the hadronic form factor. This can be achieved by multiplying $J_R-\frac{1}{2}$ dipole form factors \eq{d_HFF} with the Gaussian form factor \eq{G_HFF}. The following explicit form for the modified hadronic form factor is suggested
\begin{align}
&F_{mG}(s;m_R,\Lambda_R,\Gamma_R,J_R) \nonumber\\
&= \left(\frac{m_R^2\widetilde{\Gamma}^2_R(J_R)}{(s-m_R^2)^2+m_R^2\widetilde{\Gamma}^2_R(J_R)}\right)^{J_R-\frac{1}{2}} \nonumber\\
&\quad\,\times\exp\left(-\frac{(s-m_R^2)^2}{\Lambda_R^4}\right),\label{mG_HFF}
\end{align}
and is dubbed a ``multidipole-Gauss form factor''. In order to preserve the interpretation of the decay width of the resonance as the \textsc{fwhm} of the resonance peak, a modified decay width $\widetilde{\Gamma}_R$ is required. The explicit expression for the modified decay width depends on the spin of the resonance and reads
\begin{align}
\widetilde{\Gamma}_R(J_R) = \frac{\Gamma_R}{\sqrt{2^{\frac{1}{2J_R}}-1}}. \label{modified_decay_width}
\end{align}
The details of the derivation of the expression for $\widetilde{\Gamma}_R(J_R)$ have been diverted to Appendix \ref{sec:decay_width}. The above choice for $F_{mG}$ is inspired by the expression for the squared absolute value of the propagator denominator, i.e.\
\begin{align}
&\left(p_R^2 - m_R^2+im_R\widetilde{\Gamma}_R\right)^{-1}\left(p_R^2 - m_R^2-im_R\widetilde{\Gamma}_R\right)^{-1} \nonumber\\
&= \left((s-m_R^2)^2+m_R^2\widetilde{\Gamma}^2_R\right)^{-1},
\end{align}
where $p_R^2 = s$ has been substituted. In this way the multidipole-Gauss form factor raises the multiplicity of the propagator pole at the resonance mass.

\begin{figure}[t]
\centering
\subfigure{\includegraphics[scale=0.4]{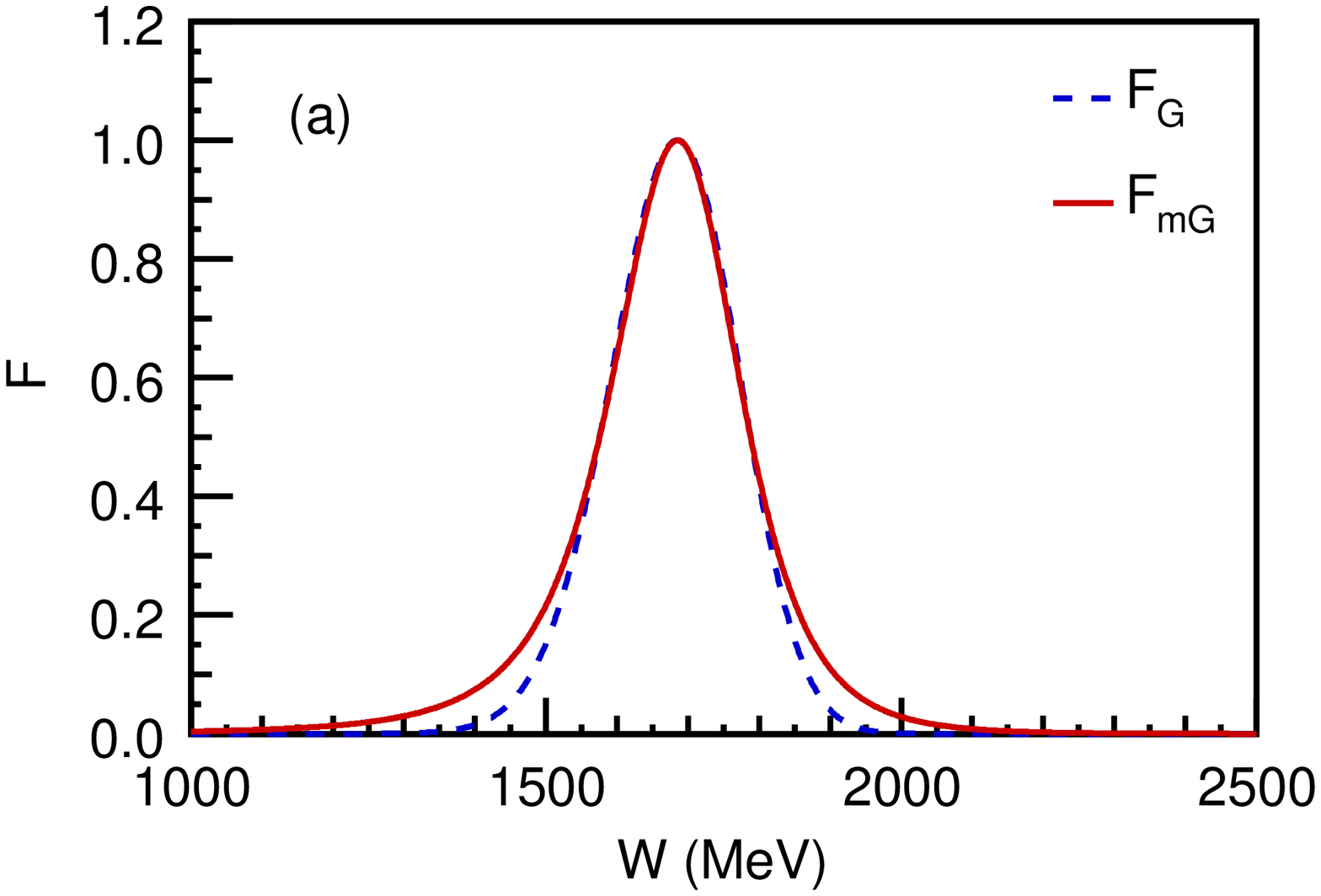}}
\subfigure{\includegraphics[scale=0.4]{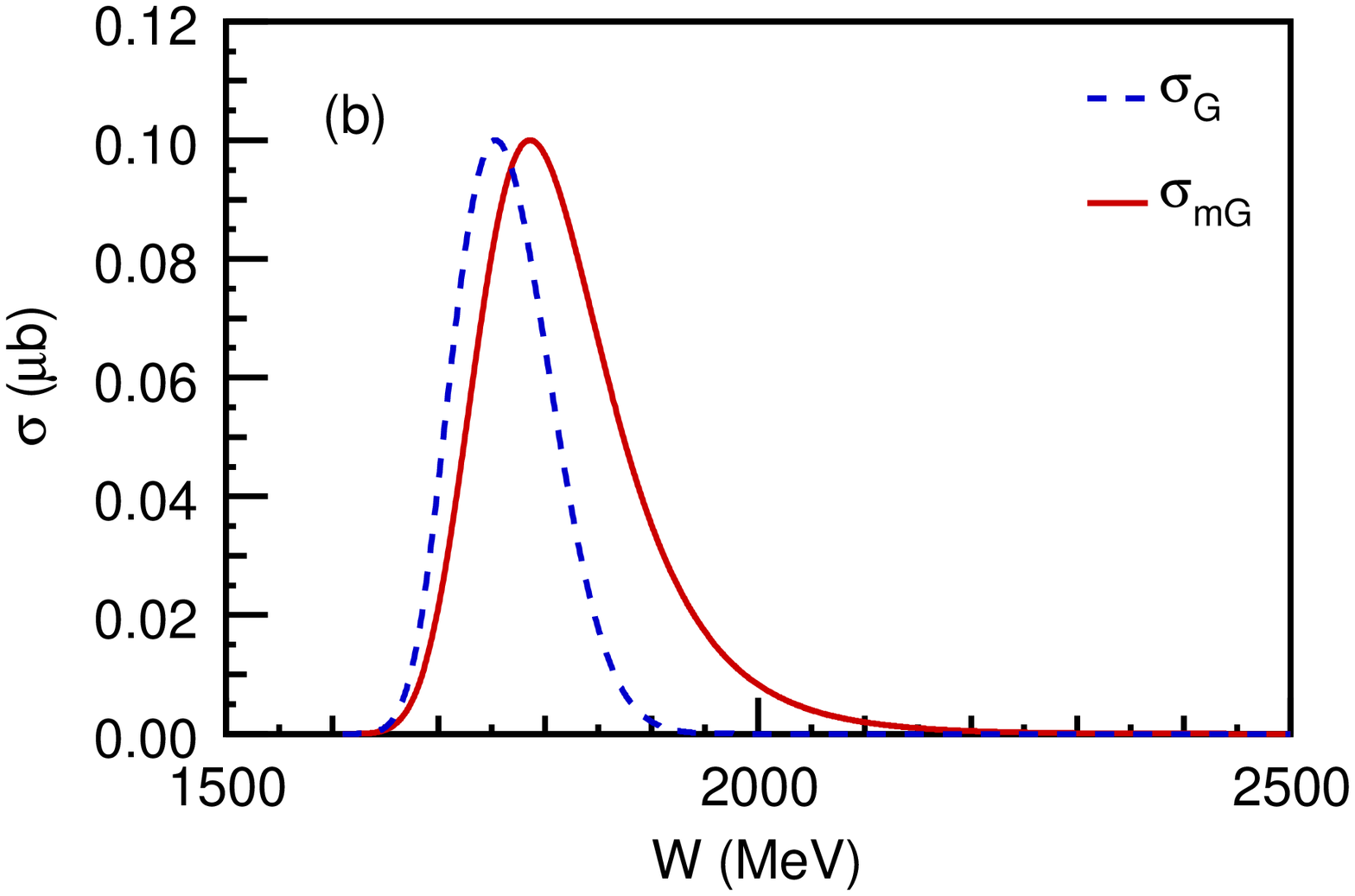}}
\caption{(a) The multidipole-Gauss form factor with $\Lambda_R = 1500$ MeV (solid curve) and the Gaussian form factor with $\Lambda_R \approx 655$ MeV (dashed curve) for the \mbox{$N(1680)\;F_{15}$} as a function of the invariant mass $W$. The cut-off energy of the Gaussian form factor was calculated so as to obtain the same \textsc{fwhm}. (b) The energy dependence of the corresponding $\gamma p \rightarrow N(1680)\;F_{15} \rightarrow K^+\Lambda$ cross section with $g =$ 6.2 (solid curve) and $g =$ 4.6 (dashed curve).}\label{spin-52_mG-HFF}
\end{figure}

Fig.~\ref{spin-52_mG-HFF}(a) compares a multidipole-Gauss form factor with $\Lambda_R = 1500$ MeV to a Gaussian form factor with the same \textsc{fwhm}.
Despite the fact that the two form factors shown in Fig.~\ref{spin-52_mG-HFF}(a) appear similar, their effect on $\sigma$ is vastly different. This is illustrated in Fig.~\ref{spin-52_mG-HFF}(b), which depicts the computed cross sections adopting both form factors. As explained before, the energy dependence of $\sigma_G$ should be interpreted as artificial. The peak position and width of the observed structure in $\sigma_G$ are determined by the value of the cut-off energy. The multidipole-Gauss form factor has a larger impact on the high-energy behavior of $\sigma$, and prevents the formation of an artificial bump. The form factor $F_{mG}$ augments the effect of the \mbox{$N(1680)\;F_{15}$}'s propagator and this makes sure that in the computed $\sigma_{mG}$ the resonance peak occurs at $W \approx m_R$. Yet, it appears as though the mass of the \mbox{$N(1680)\;F_{15}$} is shifted by approximately $100$ MeV. This ``mass shift'' is a threshold effect and is caused by the fact that $m_R - \frac{\Gamma_R}{2} \approx W_0$ and $\sigma(W_0) = 0$. The mass shift decreases with increasing resonance mass and decreasing resonance decay width.

\begin{figure}[t]
\centering
\subfigure{\includegraphics[scale=0.4]{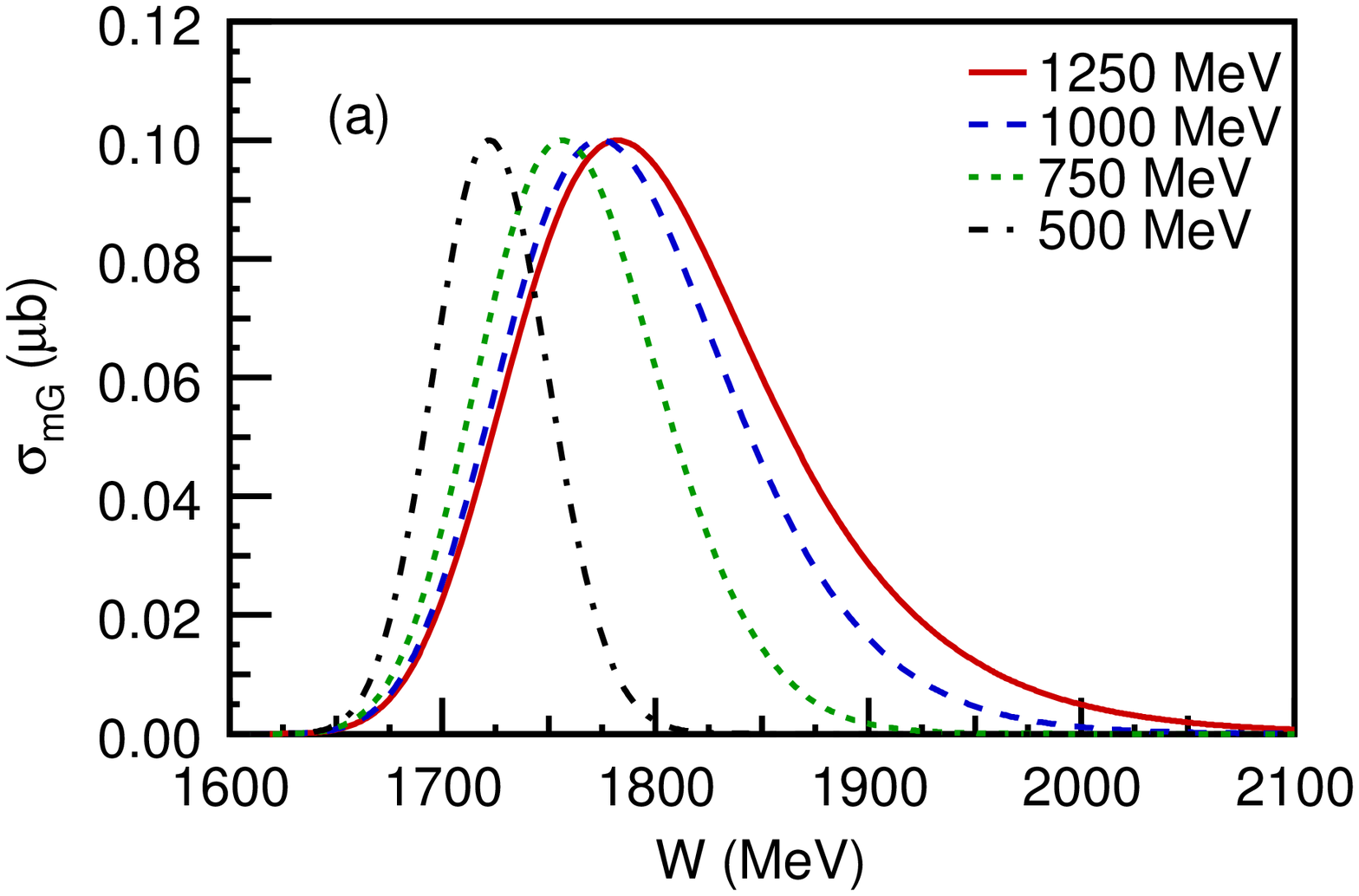}}
\subfigure{\includegraphics[scale=0.4]{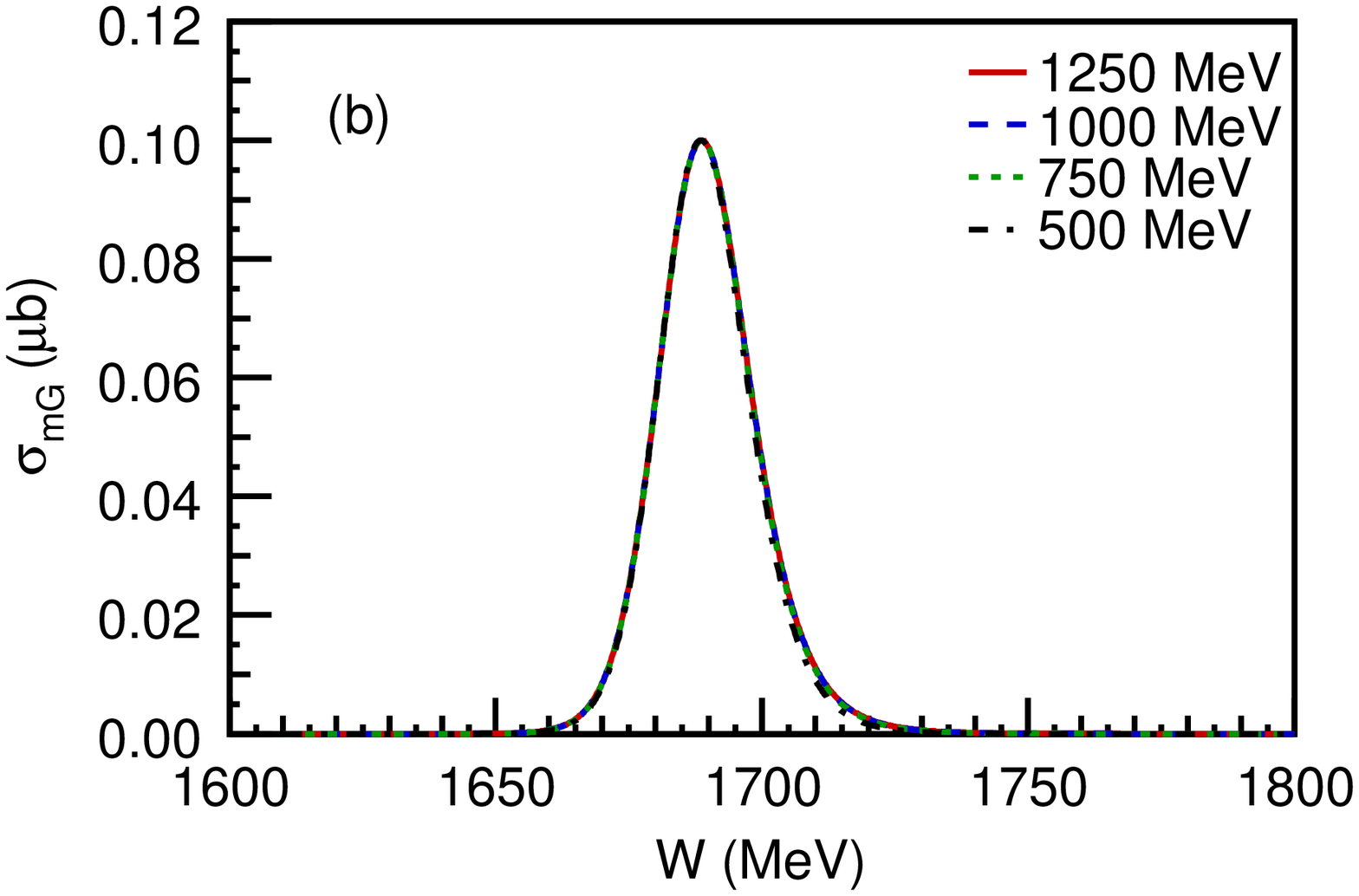}}
\caption{The energy dependence of the $\gamma p \rightarrow N(1680)\;F_{15} \rightarrow K^+\Lambda$ cross section for various values of the cut-off energy of the multidipole-Gauss form factor. In (a) the real decay width of the \mbox{$N(1680)\;F_{15}$} was used, i.e.\ $\Gamma_R = 130$ MeV, and $g =$ 6.3, 6.5, 7.0, 8.5 for $\Lambda_R =$ 1250, 1000, 750, 500 MeV. In (b) the decay width of the \mbox{$N(1680)\;F_{15}$} was set to $\Gamma_R' = 20$ MeV and $g =$ 4.2.}\label{spin-52_mG-HFF-C}
\end{figure}

In Fig.~\ref{spin-52_mG-HFF-C} the cross sections of Fig.~\ref{spin-52_bumps-shift} are replotted employing a multidipole-Gauss form factor.
The minor shift of the resonance peak in Fig.~\ref{spin-52_mG-HFF-C}(a) does not have the same origin as the artificial bump of Fig.~\ref{spin-52_bumps-shift}(a). Here the shift is caused by the comparable magnitudes of $\Lambda_R$ and $\Gamma_R$. As a consequence, the resonance peak gets increasingly narrowed by the form factor and this results in a reduction of the mass shift, which was explained in the previous paragraph. In Fig.~\ref{spin-52_bumps-shift}(b) the decay width of the \mbox{$N(1680)\;F_{15}$} was artificially lowered to $\Gamma_R' = 20$ MeV. Here it is apparent that the energy dependence of $\sigma_{mG}$ is indeed the same for all cut-off energies. Fig.~\ref{spin-52_bumps-shift}(b) also confirms the fact that the mass shift decreases with decreasing decay width: the mass shift amounts to approximately $3.5$ MeV.

\section{Conclusions}
\label{sec:conclusions}
In this work a theory of consistent interactions for massive high-spin fermions was developed. It was proven that gauge symmetry is the necessary and sufficient condition to assure the consistency of high-spin interactions. From this gauge symmetry the most general consistent interaction structure was constructed for off-shell high-spin fermions. In addition, consistent couplings for the $(\phi\psi\psi^*_{\mu_1\ldots\mu_n})$- and $(A_\mu\psi\psi^*_{\mu_1\ldots\mu_n})$-theories were derived.

It turns out that the power of the momentum dependence of consistent couplings rises with the spin of the R-S field. This gives rise to unphysical behavior in the computed tree-level cross sections if the reaction is cut off by a standard hadronic form factor. A persuasive solution was proposed in terms of an alternative phenomenological hadronic form factor, namely the multidipole-Gauss form factor. We deem that this form factor in conjunction with the consistent interaction Lagrangians developed in Sec.~\ref{subsubsec:consistent_couplings} provides a proper framework to implement the exchange of high-spin resonances in hadrodynamical analyses.

\section*{Acknowledgements}
This work was supported by the Fund for Scientific Research Flanders and the Research Council of Ghent University.

\appendix
\section{The projection operators for the spin-${5/2}$ theory}
\label{appendix:projection_operators}
The projection operators for the spin-${5/2}$ theory read \cite{Shklyar:2009cx}
\begin{align}
\P^{{5/2}}_{\mu\nu;\lambda\rho}(p) &= \frac{1}{2}\left(\P_{\mu\lambda}\P_{\nu\rho} + \P_{\mu\rho}\P_{\nu\lambda}\right) - \frac{1}{5}\P_{\mu\nu}\P_{\lambda\rho}  \nonumber\\
&\quad\,- \frac{1}{10}\bigl(\feynP{\P}_\mu\feynP{\P}_\lambda\P_{\nu\rho} + \feynP{\P}_\mu\feynP{\P}_\rho\P_{\nu\lambda} \nonumber\\
&\quad\,+ \feynP{\P}_\nu\feynP{\P}_\lambda\P_{\mu\lambda} + \feynP{\P}_\nu\feynP{\P}_\rho\P_{\mu\rho}\bigr),\label{P52}
\end{align}
\begin{align}
\P^{{3/2}}_{11,\mu\nu;\lambda\rho}(p) &= \frac{1}{2}(\P_{\mu\lambda}\mathcal{Q}_{\nu\rho} + \P_{\nu\lambda}\mathcal{Q}_{\mu\rho} \nonumber\\
&\quad\,+\P_{\mu\rho}\mathcal{Q}_{\nu\lambda} + \P_{\nu\rho}\mathcal{Q}_{\mu\lambda}) - \frac{1}{6p^2}\mathcal{R}_{\mu\nu}\mathcal{R}_{\lambda\rho}, \nonumber\\
\P^{{3/2}}_{22,\mu\nu;\lambda\rho}(p) &= \frac{1}{10}\bigl(\feynP{\P}_\mu\feynP{\P}_\lambda\P_{\nu\rho} + \feynP{\P}_\mu\feynP{\P}_\rho\P_{\nu\lambda} \nonumber\\ &\quad\,+ \feynP{\P}_\nu\feynP{\P}_\lambda\P_{\mu\lambda} + \feynP{\P}_\nu\feynP{\P}_\rho\P_{\mu\rho}\bigr) - \frac{2}{15} \P_{\mu\nu}\P_{\lambda\rho}, \nonumber\\
\P^{{3/2}}_{21,\mu\nu;\lambda\rho}(p) &= \frac{1}{2\sqrt{5}p^2} \bigl(p_\lambda\left(\feynP{\P}_\mu\P_{\nu\rho} + \feynP{\P}_\nu\P_{\mu\rho}\right) \nonumber\\
&\quad\,+ p_\rho\left(\feynP{\P}_\mu\P_{\nu\lambda} + \feynP{\P}_\nu\P_{\mu\lambda}\right)\bigr)\feynp{p} \nonumber\\
&\quad\,- \frac{1}{3\sqrt{5}p^2}\P_{\mu\nu}\mathcal{R}_{\lambda\rho}\feynp{p} \nonumber\\
&= - \P^{{3/2}}_{21,\lambda\rho;\mu\nu}(p),\label{P32}
\end{align}
\begin{align}
\P^{{1/2}}_{11,\mu\nu;\lambda\rho}(p) &= \mathcal{Q}_{\mu\nu}\mathcal{Q}_{\lambda\rho}, \nonumber\\
\P^{{1/2}}_{22,\mu\nu;\lambda\rho}(p) &= \frac{1}{3}\P_{\mu\nu}\P_{\lambda\rho}, \nonumber\\
\P^{{1/2}}_{33,\mu\nu;\lambda\rho}(p) &= \frac{1}{6p^2}\mathcal{R}_{\mu\nu}\mathcal{R}_{\lambda\rho}, \nonumber\\
\P^{{1/2}}_{21,\mu\nu;\lambda\rho}(p) &= \frac{1}{\sqrt{3}}\P_{\mu\nu}\mathcal{Q}_{\lambda\rho}, \nonumber\\
&= \P^{{1/2}}_{12,\lambda\rho;\mu\nu}(p), \nonumber\\
\P^{{1/2}}_{31,\mu\nu;\lambda\rho}(p) &= \frac{1}{\sqrt{6}p^2}\mathcal{R}_{\mu\nu}\mathcal{Q}_{\lambda\rho}\feynp{p}, \nonumber\\
&= -\P^{{1/2}}_{13,\lambda\rho;\mu\nu}(p), \nonumber\\
\P^{{1/2}}_{23,\mu\nu;\lambda\rho}(p) &= -\frac{1}{3\sqrt{2}p^2}\P_{\mu\nu}\mathcal{R}_{\lambda\rho}\feynp{p}, \nonumber\\
&= -\P^{{1/2}}_{32,\lambda\rho;\mu\nu}(p),\label{P12}
\end{align}
with
\begin{align}
\P_{\mu\nu}(p) &= g_{\mu\nu} - \frac{1}{p^2}p_\mu p_\nu, \nonumber\\
\feynP{\P}_\mu(p) &= \gamma^\nu\P_{\mu\nu}(p), \nonumber\\
&= \gamma_\mu - \frac{\feynp{p}}{p^2}p_\mu, \nonumber\\
\mathcal{Q}_{\mu\nu}(p) &= \frac{1}{p^2}p_\mu p_\nu, \nonumber\\
\mathcal{R}_{\mu\nu}(p) &= p_\mu\feynP{\P}_\nu + \feynP{\P}_\mu p_\nu, \nonumber\\
&= \left(\gamma_\mu p_\nu + \gamma_\nu p_\mu\right) - \frac{2}{p^2}\feynp{p}p_\mu p_\nu.
\end{align}
The projection operators respectively project the spin-${5/2}$ field $\psi_{\mu\nu}$ onto the (physical) spin-${5/2}$ component and the (unphysical) spin-${3/2}$ and spin-${1/2}$ components. The following orthogonality relations apply to these operators \cite{Berends:1979rv}
\begin{align}
&g^{\sigma\sigma'}g^{\tau\tau'}\P^J_{il,\mu\nu;\sigma\tau}(p)\P^{J'}_{kj,\sigma'\tau';\lambda\rho}(p) \nonumber\\
&= \delta_{JJ'}\delta_{lk}\P^{J}_{ij,\mu\nu;\lambda\rho}(p).
\end{align}

\section{The spin-$\left(n+{1/2}\right)$ projection operator}
\label{sec:simplification_projection_operator}
The $k$-th term in expression \eq{projection-general-1} for the spin-$\left(n+{1/2}\right)$ projection operator reads
\begin{align}
&\frac{n+1}{2n+3}\frac{1}{(n+1)!^2}\gamma^{\mu_1}\Biggl(\sum_{P(\mu)}\sum_{P(\nu)} A_k^{n+1} \nonumber\\
&\times\underbrace{\P_{\mu_1\mu_2}\P_{\nu_1\nu_2}\cdots\P_{\mu_{2k-1}\mu_{2k}}\P_{\nu_{2k-1}\nu_{2k}}}_{4^kk!^2} \nonumber\\
&\times\underbrace{\prod_{i=2k+1}^{n+1}\P_{\mu_i\nu_i}}_{(n-2k+1)!}\Biggr)\gamma^{\nu_1}.
\end{align}
The braces indicate how many times the relevant factor can be transformed into itself by applying permutations of the Lorentz indices, which are contained in the double sum. From now on, the factor $\frac{n+1}{2n+3}\frac{1}{(n+1)!^2}$ and the double sum are dropped for reasons of simplicity. By explicitly carrying out the contractions with the Dirac matrices, five different terms are obtained, namely
\begin{align}
&A_k^{n+1}\feynP{\P}_{\mu_2}\feynP{\P}_{\nu_2}\underbrace{\P_{\mu_3\mu_4}\P_{\nu_3\nu_4}\cdots\P_{\mu_{2k-1}\mu_{2k}}\P_{\nu_{2k-1}\nu_{2k}}}_{4^{k-1}(k-1)!^2} \nonumber\\
&\times\underbrace{\prod_{i=2k+1}^{n+1}\P_{\mu_i\nu_i}}_{(n-2k+1)!},\label{P-term-1}
\end{align}
\begin{align}
&A_k^{n+1}\underbrace{\feynP{\P}_{\mu_2}\feynP{\P}_{\mu_3}\P_{\nu_2\nu_3}\P_{\mu_4\mu_5}\P_{\nu_4\nu_5}\cdots\P_{\mu_{2k}\mu_{2k+1}}\P_{\nu_{2k}\nu_{2k+1}}}_{4^kk!(k-1)!} \nonumber\\
&\times\underbrace{\prod_{i=2k+2}^{n+1}\P_{\mu_i\nu_i}}_{(n-2k)!},\label{P-term-2}
\end{align}
\begin{align}
&A_k^{n+1}\underbrace{\P_{\mu_2\mu_3}\feynP{\P}_{\nu_2}\feynP{\P}_{\nu_3}\P_{\mu_4\mu_5}\P_{\nu_4\nu_5}\cdots\P_{\mu_{2k}\mu_{2k+1}}\P_{\nu_{2k}\nu_{2k+1}}}_{4^kk!(k-1)!} \nonumber\\
&\times\underbrace{\prod_{i=2k+2}^{n+1}\P_{\mu_i\nu_i}}_{(n-2k)!},\label{P-term-3}
\end{align}
\begin{align}
&A_k^{n+1}\left(\gamma^{\mu_1}\P_{\mu_1\nu_1}\gamma^{\nu_1}\right)\underbrace{\P_{\mu_2\mu_3}\P_{\nu_2\nu_3}\cdots\P_{\mu_{2k}\mu_{2k+1}}\P_{\nu_{2k}\nu_{2k+1}}}_{4^kk!^2} \nonumber\\
&\times\underbrace{\prod_{i=2k+2}^{n+1}\P_{\mu_i\nu_i}}_{(n-2k)!},\label{P-term-4}
\end{align}
\begin{align}
&A_k^{n+1}\feynP{\P}_{\nu_{2}}\feynP{\P}_{\mu_{2}}\underbrace{\P_{\mu_3\mu_4}\P_{\nu_3\nu_4}\cdots\P_{\mu_{2k+1}\mu_{2k+2}}\P_{\nu_{2k+1}\nu_{2k+2}}}_{4^kk!^2} \nonumber\\
&\times\underbrace{\prod_{i=2k+3}^{n+1}\P_{\mu_i\nu_i}}_{(n-2k-1)!}.\label{P-term-5}
\end{align}
Note that only expressions \eq{P-term-4} and \eq{P-term-5} have a $k = 0$ term. All of these terms have to be multiplied with a factor that accounts for the modified number of permutations that can be performed to transform the relevant term into itself. These factors read
\begin{align}
\textrm{\eq{P-term-1}} \; \rightarrow \; \frac{4^kk!^2(n-2k+1)!}{4^{k-1}(k-1)!^2(n-2k+1)!} = 4k^2,
\end{align}
\begin{align}
\textrm{\eq{P-term-2}} \; \rightarrow \; \frac{4^kk!^2(n-2k+1)!}{4^kk!(k-1)!(n-2k)!} = k(n-2k+1),
\end{align}
\begin{align}
\textrm{\eq{P-term-3}} \; \rightarrow \; \frac{4^kk!^2(n-2k+1)!}{4^kk!(k-1)!(n-2k)!} = k(n-2k+1),
\end{align}
\begin{align}
\textrm{\eq{P-term-4}} \; \rightarrow \; \frac{4^kk!^2(n-2k+1)!}{4^kk!^2(n-2k)!} = n-2k+1,
\end{align}
\begin{align}
\textrm{\eq{P-term-5}} \; \rightarrow \; \frac{4^kk!^2(n-2k+1)!}{4^kk!^2(n-2k-1)!} = (n-2k+1)(n-2k).
\end{align}
By renaming the Lorentz indices and using the following properties
\begin{align}
\feynP{\P}_\mu\feynP{\P}_\nu &= -\feynP{\P}_\nu\feynP{\P}_\mu + 2 \P_{\mu\nu}, \nonumber\\
\gamma^\mu\P_{\mu\nu}\gamma^\nu &= 3,
\end{align}
expressions \eq{P-term-1} to \eq{P-term-5} can be transformed into
\begin{align}
\textrm{\eq{P-term-1}} &\rightarrow 4(k+1)^2A_{k+1}^{n+1} \nonumber \\
&\quad\,\times\feynP{\P}_{\mu_1}\feynP{\P}_{\nu_1}\P_{\mu_2\mu_3}\P_{\nu_2\nu_3}\cdots\P_{\mu_{2k}\mu_{2k+1}}\P_{\nu_{2k}\nu_{2k+1}} \nonumber\\
&\quad\,\times\prod_{i=2k+2}^n\P_{\mu_i\nu_i},\label{new_P-term-1}
\end{align}
\begin{align}
\textrm{\eq{P-term-2}} &\rightarrow 2k(n-2k+1)A_{k}^{n+1} \nonumber \\
&\quad\,\times\P_{\mu_1\mu_2}\P_{\nu_1\nu_2}\cdots\P_{\mu_{2k-1}\mu_{2k}}\P_{\nu_{2k-1}\nu_{2k}} \nonumber \\
&\quad\,\times\prod_{i=2k+1}^n\P_{\mu_i\nu_i},\label{new_P-term-2}
\end{align}
\begin{align}
\textrm{\eq{P-term-3}} &\rightarrow 2k(n-2k+1)A_{k}^{n+1} \nonumber \\
&\quad\,\times\P_{\mu_1\mu_2}\P_{\nu_1\nu_2}\cdots\P_{\mu_{2k-1}\mu_{2k}}\P_{\nu_{2k-1}\nu_{2k}} \nonumber\\
&\quad\,\times\prod_{i=2k+1}^n\P_{\mu_i\nu_i},\label{new_P-term-3}
\end{align}
\begin{align}
\textrm{\eq{P-term-4}} &\rightarrow 3(n-2k+1)A_k^{n+1} \nonumber\\
&\quad\,\times\P_{\mu_1\mu_2}\P_{\nu_1\nu_2}\cdots\P_{\mu_{2k-1}\mu_{2k}}\P_{\nu_{2k-1}\nu_{2k}} \nonumber\\
&\quad\,\times\prod_{i=2k+1}^n\P_{\mu_i\nu_i},\label{new_P-term-4}
\end{align}
\begin{align}
\textrm{\eq{P-term-5}} &\rightarrow -(n-2k+1)(n-2k)A_k^{n+1} \nonumber\\
&\quad\,\times\feynP{\P}_{\mu_1}\feynP{\P}_{\nu_1}\P_{\mu_2\mu_3}\P_{\nu_2\nu_3}\cdots\P_{\mu_{2k}\mu_{2k+1}}\P_{\nu_{2k}\nu_{2k+1}}\nonumber\\
&\quad\,\times\prod_{i=2k+2}^n\P_{\mu_i\nu_i} \nonumber \\
&\quad\,+2(n-2k+1)(n-2k)A_k^{n+1} \nonumber\\
&\quad\,\times\P_{\mu_1\nu_1}\P_{\mu_2\mu_3}\P_{\nu_2\nu_3}\cdots\P_{\mu_{2k}\mu_{2k+1}}\P_{\nu_{2k}\nu_{2k+1}} \nonumber\\
&\quad\,\times\prod_{i=2k+2}^n\P_{\mu_i\nu_i}.\label{new_P-term-5}
\end{align}
Here, the substitution $k \rightarrow k+1$ was made for expression \eq{new_P-term-1} and the $k = 0$ term was added for expressions \eq{new_P-term-2} and \eq{new_P-term-3}, since they are proportional to $k$. In this way, all expressions obtain a $k = 0$ term. Note that the second term of expression \eq{new_P-term-5} can be rewritten as
\begin{align}
&2(n-2k+1)(n-2k)A_k^{n+1} \nonumber\\
&\P_{\mu_1\nu_1}\P_{\mu_2\mu_3}\P_{\nu_2\nu_3}\cdots\P_{\mu_{2k}\mu_{2k+1}}\P_{\nu_{2k}\nu_{2k+1}} \nonumber\\
&\prod_{i=2k+2}^n\P_{\mu_i\nu_i},\nonumber\\
&= 2(n-2k+1)(n-2k)A_k^{n+1}\nonumber\\
&\P_{\mu_1\mu_2}\P_{\nu_1\nu_2}\cdots\P_{\mu_{2k-1}\mu_{2k}}\P_{\nu_{2k-1}\nu_{2k}}\nonumber\\
&\prod_{i=2k+1}^{n-1}\P_{\mu_i\nu_i},\label{new_P-term-5a} \\
&= 2(n-2k+1)(n-2k)A_k^{n+1}\nonumber\\
&\P_{\mu_1\mu_2}\P_{\nu_1\nu_2}\cdots\P_{\mu_{2k-1}\mu_{2k}}\P_{\nu_{2k-1}\nu_{2k}}\nonumber\\
&\prod_{i=2k+1}^n\P_{\mu_i\nu_i},\label{new_P-term-5b}
\end{align}
The transition from expression \eq{new_P-term-5a} to \eq{new_P-term-5b} is valid since both expressions are equal for odd $n$ and for even $n$ the $k = \frac{n}{2}$ term, i.e.\ the last one, vanishes due to the prefactor $(n-2k)$.
Next, the factor $\frac{n+1}{2n+3}\frac{1}{(n+1)!^2}$ and the double sum are introduced again. The sum of expressions \eq{new_P-term-2}, \eq{new_P-term-3}, \eq{new_P-term-4} and \eq{new_P-term-5b} equals
\begin{align}
&\sum_{P(\mu)}\sum_{P(\nu)}\sum_{k=0}^{k_{\textrm{max},1}} \mathcal{A}_k^n\P_{\mu_1\mu_2}\P_{\nu_1\nu_2}\cdots\P_{\mu_{2k+1}\mu_{2k+2}}\P_{\nu_{2k+1}\nu_{2k+2}} \nonumber\\
&\times\prod_{i=2k+3}^n\P_{\mu_i\nu_i}.
\end{align}
The coefficients $\mathcal{A}_k^n$ can be calculated as
\begin{align}
\mathcal{A}_k^n &= \frac{n+1}{2n+3}\frac{1}{(n+1)!^2}\bigl(3(n-2k+1) + 4k(n-2k+1) \nonumber\\
&\quad\,+ 2(n-2k+1)(n-2k)\bigr)A_k^{n+1}, \nonumber\\
&= \frac{(n-2k+1)}{(n+1)!n!}A_k^{n+1}.
\end{align}
From the definition of the coefficients $A^n_k$, i.e.\ Eq.~\eq{Akn-coefficients}, the expression for $\mathcal{A}_k^n$ can be further reduced to
\begin{align}
\mathcal{A}_k^n &= \frac{(n-2k+1)}{(n+1)!n!}\biggl(\frac{1}{(-2)^k}\frac{(n+1)!}{k!(n-2k+1)!} \nonumber\\
&\quad\,\times\frac{(2n-2k+1)!!}{(2n+1)!!}\biggr), \nonumber\\
&= \frac{1}{(-2)^k}\frac{1}{n!k!(n-2k)!}\frac{(2n-2k+1)!!}{(2n+1)!!}.
\end{align}
For even values of $n$ one has $k_{\textrm{max},1} = \frac{n}{2}$. For odd values of $n$ this becomes $k_{\textrm{max},1} = \frac{n-1}{2}$.

Finally, the sum of expression \eq{new_P-term-1} and the first term of expression \eq{new_P-term-5} equals
\begin{align}
&\sum_{P(\mu)}\sum_{P(\nu)}\feynP{\P}_{\mu_1}\feynP{\P}_{\nu_1} \nonumber\\
&\times\sum_{k=0}^{k_{\textrm{max},2}} \mathcal{B}_k^n\P_{\mu_2\mu_3}\P_{\nu_2\nu_3}\cdots\P_{\mu_{2k}\mu_{2k+1}}\P_{\nu_{2k}\nu_{2k+1}}\nonumber\\
&\times\prod_{i=2k+2}^n\P_{\mu_i\nu_i}.
\end{align}
The coefficients $\mathcal{B}_k^n$ are then given by
\begin{align}
\mathcal{B}_k^n &= \frac{n+1}{2n+3}\frac{1}{(n+1)!^2}\bigl(4(k+1)^2A_{k+1}^{n+1}  \nonumber\\
&\quad\,- (n-2k+1)(n-2k)A_k^{n+1}\bigr), \nonumber\\
&= \frac{n+1}{2n+3}\frac{1}{(n+1)!^2}\biggl(4(k+1)^2 \nonumber\\
&\quad\,\times\frac{1}{(-2)^{k+1}}\frac{(n+1)!}{(k+1)!(n-2k-1)!}\frac{(2n-2k-1)!!}{(2n+1)!!} \nonumber\\
&\quad\,- (n-2k+1)(n-2k) \nonumber\\
&\quad\,\times\frac{1}{(-2)^k}\frac{(n+1)!}{k!(n-2k+1)!}\frac{(2n-2k+1)!!}{(2n+1)!!}\biggr), \nonumber\\
&= -\frac{1}{(-2)^k}\frac{1}{n!k!(n-2k-1)!}\frac{(2n-2k-1)!!}{(2n+1)!!}.
\end{align}
For even values of $n$ one has $k_{\textrm{max},1} = \frac{n-2}{2}$. For odd values of $n$ this becomes $k_{\textrm{max},1} = \frac{n-1}{2}$.

\section{The modified decay width for the multidipole-Gauss form factor}
\label{sec:decay_width}
The expression for the total cross section of Eq.~\eq{tds} is proportional to the following factor
\begin{align}
\sigma(s) \propto \left((s-m_R^2)^2+m_R^2\widetilde{\Gamma}^2_R\right)^{-2J_R},
\end{align}
which is stemming from the squared multidipole-Gauss form factor and the squared spin-$J_R$ propagator denominator. The values of $s$ corresponding to the half maximum of this factor are the solutions to the equation
\begin{align}
\left((s-m_R^2)^2+m_R^2\widetilde{\Gamma}^2_R\right)^{2J_R} = 2\left(m_R^2\widetilde{\Gamma}^2_R\right)^{2J_R}.
\end{align}
These are found as
\begin{align}
s_{\pm} = m_R^2\left(1 \pm \frac{\widetilde{\Gamma}_R}{m_R}\sqrt{2^{\frac{1}{2J_R}}-1}\right).
\end{align}
The full width at half maximum is then calculated as
\begin{align}
(\textsc{fwhm})_R &= \sqrt{s_+} - \sqrt{s_-}, \nonumber\\
&= m_R\Biggl(\sqrt{1 + \frac{\widetilde{\Gamma}_R}{m_R}\sqrt{2^{\frac{1}{2J_R}}-1}} \nonumber\\
&\quad\,- \sqrt{1 - \frac{\widetilde{\Gamma}_R}{m_R}\sqrt{2^{\frac{1}{2J_R}}-1}}\Biggr).
\end{align}
If $\widetilde{\Gamma}_R$ is defined as
\begin{align}
\widetilde{\Gamma}_R(J_R) = \frac{\Gamma_R}{\sqrt{2^{\frac{1}{2J_R}}-1}},
\end{align}
then
\begin{align}
(\textsc{fwhm})_R \approx \Gamma_R  ,
\end{align}
for $\Gamma_R \ll m_R$, which is the desired result.


\begin{thebibliography}{1}

\bibitem{Rarita:1941mf}
  W.~Rarita and J.~Schwinger,
  \href{http://link.aps.org/doi/10.1103/PhysRev.60.61}{Phys.\ Rev.\ {\bf 60}, 61 (1941)}.

\bibitem{Pascalutsa:1999zz}
  V.~Pascalutsa and R.~Timmermans,
  \href{http://link.aps.org/doi/10.1103/PhysRevC.60.042201}{Phys.\ Rev.\ C {\bf 60}, 042201 (1999)}.

\bibitem{Johnson:1960vt}
  K.~Johnson and E.~C.~G.~Sudarshan,
  \href{http://dx.doi.org/10.1016/0003-4916(61)90030-6}{Ann.\ Phys.\ {\bf 13}, 126 (1961)}.

\bibitem{Velo:1969bt}
  G.~Velo and D.~Zwanziger,
  \href{http://link.aps.org/doi/10.1103/PhysRev.186.1337}{Phys.\ Rev.\ {\bf 186}, 1337 (1969)}; \href{
http://link.aps.org/doi/10.1103/PhysRev.188.2218}{{\bf 188}, 2218 (1969)}.

\bibitem{Pascalutsa:1998pw}
  V.~Pascalutsa,
  \href{http://link.aps.org/doi/10.1103/PhysRevD.58.096002}{Phys.\ Rev.\ D {\bf 58}, 096002 (1998)}.

\bibitem{Shklyar:2009cx}
  V.~Shklyar, H.~Lenske, and U.~Mosel,
  \href{http://link.aps.org/doi/10.1103/PhysRevC.82.015203}{Phys.\ Rev.\ C {\bf 82}, 015203 (2010)}.

\bibitem{Benmerrouche:1989uc}
  M.~Benmerrouche, R.~M.~Davidson, and N.~C.~Mukhopadhyay,
  \href{http://link.aps.org/doi/10.1103/PhysRevC.39.2339}{Phys.\ Rev.\ C {\bf 39}, 2339 (1989)}.

\bibitem{Berends:1979rv}
  F.~A.~Berends, J.~W.~van Holten, P.~van Nieuwenhuizen, and B.\ De Wit,
  \href{http://dx.doi.org/10.1016/0550-3213(79)90514-5}{Nucl.\ Phys.\ B {\bf 154}, 261 (1979)}.

\bibitem{Shklyar:2004dy}
  V.~Shklyar, G.~Penner, and U.~Mosel,
  \href{http://dx.doi.org/10.1140/epja/i2004-10003-3}{Eur.\ Phys.\ J.\ A {\bf 21}, 445 (2004)}.

\bibitem{Huang:2005js}
  S.-Z.~Huang, P.-F.~Zhang, T.-N.~Ruan, Y.-C.~Zhu, and Z.-P.~Zheng,
  \href{http://dx.doi.org/10.1140/epjc/s2005-02299-4}{Eur.\ Phys.\ J.\ C {\bf 42}, 375 (2005)}.

\bibitem{Weinberg:1964cn}
  S.~Weinberg,
  \href{http://link.aps.org/doi/10.1103/PhysRev.133.B1318}{Phys.\ Rev.\ B {\bf 133}, 1318 (1964)}.

\bibitem{Behrends:1957}
  R.~E.~Behrends and C.~Fronsdal,
  \href{http://link.aps.org/doi/10.1103/PhysRev.106.345}{Phys.\ Rev.\ {\bf 106}, 345 (1957)}.

\bibitem{Janssen:2001wk}
  S.~Janssen, J.~Ryckebusch, D.~Debruyne, and T.~Van Cauteren,
  \href{http://link.aps.org/doi/10.1103/PhysRevC.65.015201}{Phys.\ Rev.\ C {\bf 65}, 015201 (2002)}.

\bibitem{Nakamura:2010zzi}
  K.~Nakamura \ea [Particle Data Group Collaboration],
  \href{http://dx.doi.org/10.1088/0954-3899/37/7A/075021}{J.\ Phys.\ G {\bf 37}, 075021 (2010)}.

\bibitem{Klempt:2009pi}
  E.~Klempt and J.~-M.~Richard,
  \href{http://dx.doi.org/10.1103/RevModPhys.82.1095}{Rev.\ Mod.\ Phys.\  {\bf 82}, 1095 (2010)}.

\bibitem{Janssen:2001pe}
  S.~Janssen, J.~Ryckebusch, W.~Van Nespen, D.~Debruyne, and T.~Van Cauteren,
  \href{http://dx.doi.org/10.1007/s100500170100}{Eur.\ Phys.\ J.\ A {\bf 11}, 105 (2001)}.

\bibitem{Benmerrouche:1994uc}
  M.~Benmerrouche, N.~C.~Mukhopadhyay, and J.~F.~Zhang,
  \href{http://link.aps.org/doi/10.1103/PhysRevD.51.3237}{Phys.\ Rev.\ D {\bf 51}, 3237 (1995)}.

\bibitem{Pearce:1990uj}
  B.~C.~Pearce and B.~K.~Jennings,
  \href{http://dx.doi.org/10.1016/0375-9474(91)90256-6}{Nucl.\ Phys.\ A {\bf 528}, 655 (1991)}.

\bibitem{Corthals:2005ce}
  T.~Corthals, J.~Ryckebusch, and T.~Van Cauteren,
  \href{http://link.aps.org/doi/10.1103/PhysRevC.73.045207}{Phys.\ Rev.\ C {\bf 73}, 045207 (2006)}.

\end{thebibliography}
\end{document}